\documentclass[12pt]{article}

\usepackage[dvips]{graphicx}
\usepackage{epsfig}
\usepackage{amsmath}
\usepackage{verbatim}
\usepackage{psfrag}
\usepackage{bm}
\usepackage{bbm}
\usepackage{color}

\usepackage{epsf,epsfig}
\usepackage{graphics}

\def\be{\begin{equation}}
\def\ee{\end{equation}}
\def\bea{\begin{eqnarray}}
\def\eea{\end{eqnarray}}

\allowdisplaybreaks

\begin{document}
\thispagestyle{empty}

\begin{flushright}
{
\small
TTK-10-16\\
ZU-TH 02/10\\
1002.1326 [hep-ph]\\[0.2cm]
February 5, 2010
}
\end{flushright}

\vspace{0.4cm}
\begin{center}
\Large\bf\boldmath
Finite Number Density Corrections to Leptogenesis
\unboldmath
\end{center}

\vspace{0.4cm}

\begin{center}
{Martin Beneke,
Bj\"orn~Garbrecht,
Matti Herranen}\\
\vskip0.3cm
{\it Institut f\"ur Theoretische Teilchenphysik 
und Kosmologie,\\ 
RWTH Aachen University,\\
D--52056 Aachen, Germany}\\
\vskip.3cm
and\\
\vskip.3cm
{Pedro Schwaller}\\
\vskip0.3cm
{\it Institut f\"ur Theoretische Physik,\\
 Universit\"at Z\"urich, CH--8057 Z\"urich, 
Switzerland}
\end{center}

\vspace{1.3cm}
\begin{abstract}
\vspace{0.2cm}\noindent
We derive and solve kinetic equations for leptogenesis within the 
Closed Time Path (CTP) formalism. It is particularly emphasised how 
the procedure of real intermediate state subtraction familiar from the 
Boltzmann approach is realised within the CTP framework; and we show 
how in time-independent situations no lepton asymmetry emerges,
in accordance with the $CPT$ theorem. The CTP approach provides new 
quantum statistical corrections from evaluating the loop integrals. 
These lead to an enhancement of the asymmetry originating from the 
Bose statistics of the Higgs particles. To quantify this effect,
we define and evaluate an effective $CP$-violating parameter. We also solve
the kinetic equations and show explicitly that the new quantum statistical
corrections can be neglected in the strong washout regime, while, depending
on initial conditions, they can be sizable for weak washout.
\end{abstract}

\newpage
\setcounter{page}{1}

\newpage
\allowdisplaybreaks[2]

\section{Introduction}

In the light of the present knowledge of elementary particle physics 
phenomenology, leptogenesis appears as one of the most plausible 
explanations for the dynamical emergence of the baryon asymmetry of 
the Universe~\cite{Fukugita:1986hr,Davidson:2008bu}. Leptogenesis in 
turn is a variant of baryogenesis from out-of-equilibrium decays of 
heavy particles during the expansion and cooling of the early 
Universe~\cite{Nanopoulos:1979gx,Kolb:1979qa}. The deviation from
equilibrium provides the breaking of time reversal invariance, that 
is a necessary condition for baryogenesis~\cite{Sakharov:1967dj} as a 
consequence of the $CPT$ theorem.

The most common approach for making quantitative predictions is to 
formulate and solve a network of Boltzmann equations, that describes 
the evolution of the number densities of the particle species relevant 
to baryogenesis. In simple scenarios for
leptogenesis, a setup that leads to meaningful results is to compute
the distribution function of a heavy right-handed Majorana neutrino $N_1$
and of the charge density of a lepton doublet $\ell$, while one may 
assume that the remaining species are in thermal equilibrium. More 
realistic models typically take account of the presence of the several 
generations. The Boltzmann equations are usually written
in a form, where on the left hand side, there is a kinetic term that traces
the particle number distributions, while on the right hand side, 
there is a collision term. The collision term accounts for elastic and 
inelastic (particle number violating) scatterings. It arises from 
weighting reaction rates, computed in vacuum scattering theory (in-out 
formalism), with the distribution functions of the species involved.

When setting up the Boltzmann equations, one encounters the following issue:
Let $\Gamma_{Ni\to \ell \phi}$, the rate for the decay process 
$N_i\to \ell \phi$, be proportional to $1+\varepsilon$, where 
$\varepsilon$ parametrises the $CP$ asymmetry of the
decay. Then, by complex conjugation of the involved coupling constants, 
it follows that $\Gamma_{Ni\to \bar\ell \phi^*}\propto 1-\varepsilon$. 
Application of the $CPT$ theorem then implies that the inverse decay rates
$\Gamma_{\ell \phi \to Ni}\propto 1-\varepsilon$
and $\Gamma_{\bar\ell \phi^*\to Ni}\propto 1+\varepsilon$. If these were 
the only contributions that are accounted for in the collision term, 
one might conclude that the two-by-two rate $\ell\phi\to\bar\ell\phi^*$ 
is proportional to $1-2\varepsilon$, while the rate of  
$\bar\ell\phi^*\to\ell\phi$ is proportional
to $1+2\varepsilon$. An asymmetry even in equilibrium would result, in apparent
contradiction with $CPT$ invariance. But of course, there is no contradiction,
since the heavy neutrinos  are unstable and therefore do not belong to
the Hilbert space of asymptotic states of a unitary $S$-matrix. The total
lepton number violating two-by-two rates are therefore not given by the
above naive multiplication of the rates for decay and inverse decay.
Yet, it is crucial that one tracks the neutrino densities in the 
Boltzmann equations. The issue is fixed when realising that the 
resonant parts of the two-by-two scatterings, where the intermediate 
neutrino is on-shell, contribute to the Boltzmann equations at
the same order in coupling constants as the decay and inverse decay processes.
Technically, one adds the full two-by-two rates and subtracts from these
the on-shell contribution, since this is already accounted for in the 
decays and inverse decays~\cite{Kolb:1979qa}. This procedure
is known as the subtraction of real intermediate states (RIS).

One expects that the RIS subtraction is naturally realised
within a first principle approach, that is less heuristic than the 
Boltzmann ansatz. This first principle approach is
given by applying the Schwinger-Keldysh Closed Time Path (CTP) 
formalism~\cite{Schwinger:1960qe,Keldysh:1964ud}
to field theory~\cite{Calzetta:1986cq}.
In this paper, we derive and solve the Kadanoff-Baym equations for 
leptogenesis, that result from the CTP formalism. We pay particular 
attention to demonstrating that no lepton asymmetry is generated in 
equilibrium, or, more generally
in situations that are time-reversal invariant. For models with
fermionic leptons this result has been reported earlier in 
works based on the CTP 
formalism~\cite{Buchmuller:2000nd,De Simone:2007rw,Anisimov:2010aq},
though no detailed derivation has been given. (The main focus of
Refs.~\cite{De Simone:2007rw,Anisimov:2010aq} is the study of memory  
effects.) For the case of scalar leptons, the vanishing of the 
asymmetry in equilibrium has been demonstrated recently in 
Refs.~\cite{Garny:2009rv,Garny:2009qn}.

In addition to offering the framework for
a first-principle derivation of equations that 
describe leptogenesis, the CTP formalism also provides finite density 
corrections that do not straightforwardly follow in the Boltzmann approach.
These corrections take the form of an integral over terms linear in 
the Higgs boson and lepton distribution functions.
Historically, it has been proposed to compute the $CP$ asymmetry 
from the discontinuity of the correlation
functions in the Matsubara (imaginary time) formalism. In the CTP (real time)
approach, this corresponds to the imaginary part of
the time-ordered correlation functions~\cite{Covi:1997dr,Giudice:2003jh}.
It is characteristic of these Green functions, that they feature not only 
terms that are linear in the lepton and Higgs boson distribution functions,
but also a product term of these distributions. In contrast, it appears 
to be the consensus among the more recent 
papers~\cite{De Simone:2007rw,Anisimov:2010aq,Garny:2009rv,Garny:2009qn} 
that the $CP$ asymmetry does not receive corrections from a product 
term of Higgs boson and lepton distribution functions. This has been 
shown in detail for the model with scalar leptons in
Refs.~\cite{Garny:2009rv,Garny:2009qn}. Here we find the same property
in the model with fermionic leptons: The $CP$ asymmetry results from 
terms that can be written as the difference of Wightman type correlators 
(no explicit time-ordering of the field operators).

After deriving the kinetic equation for the lepton asymmetry  in the 
CTP formalism we solve it numerically in various situations. We find 
that the effect of the new quantum statistical corrections is small when 
the bulk of the lepton asymmetry is produced at temperatures $T\ll M_1$, 
where $M_1$ is the mass of the neutrino that decays out of equilibrium, 
because the distribution functions are strongly Boltzmann suppressed for 
momenta much large than $T$. This situation corresponds to the 
strong-washout scenario, which is of particular interest since the 
resulting asymmetry is independent of both, the initial abundance of 
heavy neutrinos and a possible primordial 
lepton asymmetry. On the other hand, when the bulk of the asymmetry is produced for
$T\sim M_1$ or larger, which corresponds to weak washout,
the impact is more sizable. For a vanishing initial lepton asymmetry and 
thermal initial conditions of $N_1$, we find an enhancement of the 
asymmetry, as apparently the Bose enhancement from the Higgs particle 
distribution function dominates over the Fermi suppression
from the lepton distribution. This qualitatively confirms what
has been reported for scalar leptons~\cite{Garny:2009rv,Garny:2009qn}. 
In this context, we note that the effects of finite density have been 
considered when the quantum statistical factors are accounted for in the 
initial and final states~\cite{Basboll:2006yx,HahnWoernle:2009qn}, 
but not in the loop corrections, in which case it has also been found 
that the corrections may be sizable in the weak-washout regime.
However, the quantum statistical factors in the loop corrections 
appear within the CTP approach on the same footing and with the same 
size as the external state factors, which
is why they cannot be neglected consistently.

The remainder of this paper is organised as follows: In
Section~\ref{section:CTP}, we specify the model, introduce the free 
CTP propagators and provide the one-loop self energies. These ingredients 
are used in Section~\ref{Section:Tree:Level} to derive the Kadanoff-Baym
equations that correspond to the tree-level Boltzmann equations with 
quantum statistical factors. The wave-function correction that leads to a 
$CP$ asymmetry arises from a two-loop diagram in the CTP formalism, which 
we discuss in Section~\ref{section:wf}. We explicitly show that no 
asymmetry is produced in equilibrium and we relate this fact to the usual 
procedure of RIS subtraction. Similarly, we discuss the
vertex correction in Section~\ref{section:vertex}. In
Section~\ref{section:numerics}, we numerically evaluate the effective 
$CP$-violating parameter and solve the Boltzmann equations for different 
strengths of washout, which demonstrates that finite density corrections 
are important for weak washout, but may be neglected in the 
strong-washout domain. We conclude  in
Section~\ref{section:conclusions}.

\section{CTP Approach to Leptogenesis}
\label{section:CTP}

In the simplest scenario of leptogenesis, there are two right handed
neutrinos $N_i$ ($i=1,2$) with lepton-number violating Majorana masses $M_i$,
one scalar SU(2) Higgs doublet $\phi$ and an active left-handed lepton 
doublet $\ell$. The Lagrangian of the model is
\begin{align}
\label{Lagrangian}
{\cal L}=\frac{1}{2}\bar\psi_{Ni}({\rm i} \partial\!\!\!/-M_i) \psi_{Ni}
+\bar\psi_\ell{\rm i}\partial\!\!\!/\psi_\ell
+(\partial^\mu\phi^\dagger)(\partial_\mu \phi)
-Y_i^*\bar\psi_\ell \phi^\dagger P_{\rm R}\psi_{Ni}
-Y_i\bar\psi_{Ni}P_{\rm L}\phi\psi_\ell\,,
\end{align}
where a summation over $i=1,2$ is understood and 
$P_{\rm L/R}=(1\mp\gamma_5)/2$. Besides, it is implicitly assumed
that the ${\rm SU}(2)$ indices are contracted in a gauge-invariant way, 
that is $\phi\psi_\ell=-\phi_A\varepsilon_{AB}\psi_{\ell B}$
and $\bar\psi_{\ell}\phi^\dagger=(\bar\psi_{\ell})_A \varepsilon_{AB}
(\phi^\dagger)_B$, where $\epsilon_{AB}$ denotes the antisymmetric $2\times 2$ 
matrix with $\epsilon_{12}=1$. Upon the four-component spinor $\psi_{Ni}$, 
we impose the Majorana constraint
\begin{align}
\label{Majorana:condition}
\psi_{Ni}^c=C \bar\psi_{Ni}^T=\psi_{Ni}\,,
\end{align}
with $C$ the charge conjugation matrix. 

\subsection{Propagators}

For the CTP formalism we follow closely the notational conventions of 
Refs.~\cite{Prokopec:2003pj,Prokopec:2004ic,Garbrecht:2008cb} and  
define the CTP propagators for the leptons by 
\begin{subequations}
\begin{align}
\label{prop+-}
{\rm i}S_{\ell\alpha\beta}^{+-}(u,v)&={\rm i}S_{\ell\alpha\beta}^{<}(u,v)
=-\langle \bar\psi_{\ell \beta}(v) \psi_{\ell \alpha}(u)\rangle\,,\\
{\rm i}S_{\ell\alpha\beta}^{-+}(u,v)&={\rm i}S_{\ell\alpha\beta}^{>}(u,v)
=\langle\psi_{\ell \alpha}(u) \bar\psi_{\ell \beta}(v)\rangle\,,\\
{\rm i}S_{\ell\alpha\beta}^{++}(u,v)&={\rm i}S_{\ell\alpha\beta}^{T}(u,v)
=\langle T(\psi_{\ell \alpha}(u) \bar\psi_{\ell \beta}(v))\rangle\,,\\
{\rm i}S_{\ell\alpha\beta}^{--}(u,v)&={\rm i}S_{\ell\alpha\beta}^{\bar T}(u,v)
=\langle \bar T(\psi_{\ell \alpha}(u) \bar\psi_{\ell \beta}(v))\rangle
\,.
\label{prop--}
\end{align}
\end{subequations}
We perform a perturbation and gradient expansion of the full Kadanoff-Baym 
equations that uses the tree-level propagators and vertices as elementary 
building blocks. The tree-level propagators are therefore given by the 
vacuum propagators plus a finite-density contribution 
proportional to the spectral function of the free theory. When there are 
no correlations between modes of different momentum and spin,
we can characterize the state by particle number distribution functions.
In case there is also no preferred direction for the spin, the usual 
averaging procedure applies. These assumptions lead to the
explicit expressions
\begin{subequations}
\label{prop:ell:expl}
\begin{align}
{\rm i}S_{\ell}^{<}(p)
&=-2\pi\delta(p^2)P_{\rm L}p\!\!\!/P_{\rm R}\left[
\vartheta(p_0)f_{\ell}(\mathbf{p})
-\vartheta(-p_0)(1-\bar f_{\ell}(-\mathbf{p}))
\right]\,,\\
{\rm i}S_{\ell}^{>}(p)
&=-2\pi\delta(p^2)P_{\rm L}p\!\!\!/P_{\rm R}\left[
-\vartheta(p_0)(1-f_{\ell}(\mathbf{p}))
+\vartheta(-p_0)\bar f_{\ell}(-\mathbf{p})
\right]\,,\\
{\rm i}S_{\ell}^{T}(p)
&=
P_{\rm L}\frac{{\rm i}p\!\!\!/}{p^2+{\rm i}\varepsilon}P_{\rm R}
-2\pi\delta(p^2)P_{\rm L}p\!\!\!/P_{\rm R}\left[
\vartheta(p_0)f_{\ell}(\mathbf{p})
+\vartheta(-p_0)\bar f_{\ell}(-\mathbf{p})
\right]\,,\\
{\rm i}S_{\ell}^{\bar T}(p)
&=
-P_{\rm L}\frac{{\rm i}p\!\!\!/}{p^2-{\rm i}\varepsilon}P_{\rm R}
-2\pi\delta(p^2)P_{\rm L}p\!\!\!/P_{\rm R}\left[
\vartheta(p_0)f_{\ell}(\mathbf{p})
+\vartheta(-p_0)\bar f_{\ell}(-\mathbf{p})
\right]
\,,
\end{align}
\end{subequations}
for the lepton propagators, 
where $f_\ell(\mathbf{k})$ denotes the distribution function of leptons and
$\bar f_\ell(\mathbf{k})$ of anti-leptons.
The leptons $\ell$ occur within an ${\rm SU}(2)$ doublet, but since 
this symmetry is unbroken at the time of leptogenesis, we assume that 
the lepton densities are
${\rm SU}(2)$-symmetric and hence diagonal.
In particular, the lepton propagators defined here are the diagonal
components of a diagonal
propagator with additional ${\rm SU}(2)$ indices:
\begin{align}
S^{{\rm SU}(2)}_{\ell  AB}(u,v)=\delta_{AB} S_{\ell}(u,v)\,\quad A,B=1,2\,.
\end{align}
Similarly, for the Majorana-fermion propagators we find
\begin{subequations}
\label{prop:N:expl}
\begin{align}
{\rm i}S_{Ni}^{<}(p)
&=-2\pi\delta(p^2-M_i^2)(p\!\!\!/+M_i)\left[
\vartheta(p_0)f_{Ni}(\mathbf{p})
-\vartheta(-p_0)(1-f_{Ni}(-\mathbf{p}))
\right]\,,\\
{\rm i}S_{Ni}^{>}(p)
&=-2\pi\delta(p^2-M_i^2)(p\!\!\!/+M_i)\left[
-\vartheta(p_0)(1-f_{Ni}(\mathbf{p}))
+\vartheta(-p_0)f_{Ni}(-\mathbf{p})
\right]\,,\\
{\rm i}S_{Ni}^{T}(p)
&=
\frac{{\rm i}(p\!\!\!/+M_i)}{p^2-M_i^2+{\rm i}\varepsilon}
-2\pi\delta(p^2-M_i^2)(p\!\!\!/+M_i)\left[
\vartheta(p_0)f_{Ni}(\mathbf{p})
+\vartheta(-p_0)f_{Ni}(-\mathbf{p})
\right]\,,\\
{\rm i}S_{Ni}^{\bar T}(p)
&=
-\frac{{\rm i}(p\!\!\!/+M_i)}{p^2-M_i^2-{\rm i}\varepsilon}
-2\pi\delta(p^2-M_i^2)(p\!\!\!/+M_i)\left[
\vartheta(p_0)f_{Ni}(\mathbf{p})
+\vartheta(-p_0)f_{Ni}(-\mathbf{p})
\right]
\,.
\end{align}
\end{subequations}
As a consequence of the Majorana condition~(\ref{Majorana:condition}),
the distribution functions for neutrinos and anti-neutrinos are identical,
and the propagators inherit the property
\begin{align}
\label{Majorana:propagator}
S_{Ni}(u,v)=C S_{Ni}^t(v,u) C^\dagger\,,
\end{align}
where the transposition acts here on both the CTP and the Dirac indices.
Finally the scalar propagators take the form
\begin{subequations}
\label{prop:phi:expl}
\begin{align}
{\rm i}\Delta_\phi^<(p)&=
2\pi \delta(p^2)\left[
\vartheta(p_0) f_\phi(\mathbf p)
+\vartheta(-p_0) (1+\bar f_\phi(-\mathbf p))\right]
\,,
\\
{\rm i}\Delta_\phi^>(p)&=
2\pi \delta(p^2)\left[
\vartheta(p_0) (1+f_\phi(\mathbf p))
+\vartheta(-p_0) \bar f_\phi(-\mathbf p)\right]
\,,
\\
{\rm i}\Delta_\phi^T(p)&=
\frac{\rm i}{p^2+{\rm i}\varepsilon}+
2\pi \delta(p^2)\left[
\vartheta(p_0) f_\phi(\mathbf p)
+\vartheta(-p_0) \bar f_\phi(-\mathbf p)\right]
\,,
\\
{\rm i}\Delta_\phi^{\bar T}(p)&=
-\frac{\rm i}{p^2-{\rm i}\varepsilon}+
2\pi \delta(p^2)\left[
\vartheta(p_0) f_\phi(\mathbf p)
+\vartheta(-p_0) \bar f_\phi(-\mathbf p)\right]
\,.
\end{align}
\end{subequations}
As for the leptons, it is understood that
\begin{align}
\Delta^{{\rm SU}(2)}_{\phi  AB}(u,v)=\delta_{AB} \Delta_{\phi}(u,v)\,
\quad A,B=1,2\,.
\end{align}
In the following, an occasional multiplicity factor $g_w=2$ arises from
performing ${\rm SU}(2)$ traces such as
\begin{align}
{\rm tr}_{{\rm SU}(2)}[\Delta_\phi^{{\rm SU}(2)} \epsilon^\dagger 
S_\ell^{{\rm SU}(2)} \epsilon]
= \Delta_{\phi} S_{\ell}\,{\rm tr}_{{\rm SU}(2)}[\epsilon^\dagger 
\epsilon]
=g_w \Delta_{\phi} S_{\ell}\,.
\end{align}
In thermal equilibrium the distribution functions assume the usual 
Fermi-Dirac and Bose-Einstein forms, respectively:
\begin{align}\label{eqdists}
f_\ell(\mathbf p)=\bar f_\ell(\mathbf p)=f^{\rm eq}_\ell(\mathbf p)
&=\frac{1}{e^{\beta |\mathbf p|}+1}\,,\notag\\
f_{Ni}(\mathbf p)=f^{\rm eq}_{Ni}(\mathbf p)
&=\frac{1}{e^{\beta \sqrt{\mathbf p^2+M_i^2}}+1}\,, \\
f_\phi(\mathbf p)=\bar f_\phi(\mathbf p)=f^{\rm eq}_\phi(\mathbf p)
&=\frac{1}{e^{\beta |\mathbf p|}-1}\notag\,.
\end{align}
Furthermore the equilibrium Green functions satisfy the 
Kubo-Martin-Schwinger (KMS) relation,
\begin{align}
\label{KMS}
{\rm i}S_\ell^>(p)=-{\rm e}^{\beta p_0}\,{\rm i}S_\ell^<(p)\,,
\quad
{\rm i}S_{Ni}^>(p)=-{\rm e}^{\beta p_0}\,{\rm i}S_{Ni}^<(p)\,,
\quad
{\rm i}\Delta_\phi^>(p)={\rm e}^{\beta p_0}\,{\rm i}\Delta_\phi^<(p)\,,
\end{align}
an extremely useful property that we employ exhaustively throughout this
paper.

\subsection{One-loop self-energies}

The one-loop neutrino self-energy in the CTP formalism  is given by 
({\it cf.} the vacuum expression in Ref.~\cite{Pilaftsis:1997jf})
\begin{align}
\label{Sigma:N}
{\rm i}{\Sigma\!\!\!/}^{ab}_{Nij}(k)
=& \,g_w \int\frac{d^4 k^\prime}{(2\pi)^4}\frac{d^4 k^{\prime\prime}}{(2\pi)^4}
(2\pi)^4 \delta^{(4)}(k-k^\prime-k^{\prime\prime})\\
\notag
&\times
\Big\{
Y_i Y_j^* P_{\rm L} {\rm i}S_{\ell}^{ab}(k^\prime) P_{\rm R} 
{\rm i} \Delta_\phi^{ab}(k^{\prime\prime})
+Y_i^* Y_j C \left[P_{\rm L} {\rm i}S_{\ell}^{ba} (-k^\prime)
P_{\rm R}\right]^T C^\dagger
{\rm i} \Delta_\phi^{ba}(-k^{\prime\prime})
\Big\}\,,
\end{align}
where $a,b=\pm$ denote the CTP indices, see 
Eqs.~(\ref{prop+-})--(\ref{prop--}). Our definition for the self-energies 
is such that $-i\Sigma$ corresponds to the sum of all one-particle 
irreducible diagrams. The one-loop lepton self-energy reads
\begin{align}
\label{Sigma:ell}
{\rm i}{\Sigma\!\!\!\!/}_\ell^{ab}(k)= |Y_i|^2
\int \frac{d^4 k^\prime}{(2\pi)^4}\frac{d^4 k^{\prime\prime}}{(2\pi)^4}
(2\pi)^4 \delta^{(4)}(k-k^\prime-k^{\prime\prime})
P_{\rm R}
{\rm i}S_{Ni}^{ab}(k^\prime) 
P_{\rm L}
{\rm i}\Delta_{\phi}^{ba}(-k^{\prime\prime})
\,,
\end{align}
where we sum over the neutrino flavours $i$.
In thermal equilibrium these self-energies inherit the
KMS property
\begin{align}
\label{KMS:Sigma}
{\Sigma\!\!\!/}_{Ni,\ell}^>(p)=-{\rm e}^{\beta p_0}
\,{\Sigma\!\!\!/}^<_{Ni,\ell}(p)\,.
\end{align}
from the constituent CTP Green functions. At the one-loop order it is 
easy to verify this explicitly by inserting the tree-level
propagators~(\ref{prop:ell:expl}), (\ref{prop:N:expl}), 
(\ref{prop:phi:expl}) into the self-energies~(\ref{Sigma:N}), 
(\ref{Sigma:ell}). One of the main
objectives of this paper is to show explicitly how KMS also remains
valid when considering the next-to-leading order contributions to
${\Sigma\!\!\!/}_{Ni,\ell}^{<,>}$.  It turns out that this
encompasses the subtraction of RIS as it is performed in the conventional
approach to baryo- and leptogenesis.

\section{Tree-Level Contributions to the Kinetic Equations}
\label{Section:Tree:Level}

The theoretical description of leptogenesis is concerned with the
time evolution of non-equilibrium densities of active leptons and of
right-handed neutrinos. This is described by the Kadanoff-Baym equations
(see {\it e.g.}~\cite{Prokopec:2003pj,Prokopec:2004ic,Garbrecht:2008cb}). 
We perform the standard Wigner transformation and gradient expansion. 
Since we are interested in finite-density effects from loops in the 
CTP formalism, we restrict ourselves to the leading term in the gradient 
expansion. We regard this as a self-consistent leading-order 
approximation to the full Kadanoff-Baym equations, relative to which 
corrections from higher gradients and loops may eventually be  
considered. Thus, the kinetic equations take the form
\begin{subequations}
\begin{align} 
\label{leptoneq1}
\frac{d}{dt} \gamma_0
{\rm i}S_\ell^{<,>}(k)
&=-\left[
{\rm i}{\Sigma\!\!\!/}_\ell^{>}(k)P_{\rm L} {\rm i}S_\ell^{<}(k)
-{\rm i}{\Sigma\!\!\!/}_\ell^{<}(k)P_{\rm L} {\rm i}S_\ell^{>}(k)
\right]\,,
\\
\frac{d}{dt} \gamma_0
{\rm i}S_{Ni}^{<,>}(k)
&=-\left[
{\rm i}{\Sigma\!\!\!/}_{Nii}^{>}(k) {\rm i}S_{Ni}^{<}(k)
-{\rm i}{\Sigma\!\!\!/}_{Nii}^{<}(k) {\rm i}S_{Ni}^{>}(k)
\right]
\,,
\end{align}
\end{subequations}
where we recognize the familiar expression for the collision term in the 
CTP formalism on the right-hand side. Here and in the following we assume 
spatial isotropy, which means that there is no angular
dependence in the distribution functions and we may identify
$f_{\ell,\phi,Ni}(\mathbf k)\equiv f_{\ell,\phi,Ni}(|\mathbf k|)$.  
In particular, we repeatedly use $f_{\ell,\phi,Ni}(-\mathbf k) = 
f_{\ell,\phi,Ni}(\mathbf k)$ to simplify the arguments of distribution 
functions. When performing the Wigner transformation and gradient expansion 
in a slowly varying background, all time dependence in the
propagators~(\ref{prop:ell:expl}), (\ref{prop:N:expl}), (\ref{prop:phi:expl})
is isolated in the distribution functions $f_{\ell,\phi,Ni}$, where
we suppress an explicit time argument.
The equation for the lepton density can be simplified when
taking the Dirac trace of Eq.~(\ref{leptoneq1}), 
multiplying by minus one and integrating over $k_0$.
Inserting
Eq.~(\ref{prop:ell:expl}), we obtain
\begin{align}
\label{KB:ell}
\frac{d}{d t}
\left(f_{\ell}(\mathbf k)-\bar f_{\ell}(\mathbf k) \right)
={\cal C}_\ell(\mathbf k)
=\int \frac{d k_0}{2\pi}{\rm tr}\left[
{\rm i}{\Sigma\!\!\!/}_\ell^{>}(k)P_{\rm L} {\rm i}S_\ell^{<}(k)
-{\rm i}{\Sigma\!\!\!/}_\ell^{<}(k)P_{\rm L} {\rm i}S_\ell^{>}(k)
\right]\,,
\end{align}
which defines ${\cal C}_\ell(\mathbf k)$.
In a similar way we obtain the equation for the time dependence 
of the Majorana neutrino distribution function 
\begin{align}
\label{KB:N}
\frac{d}{dt}
f_{Ni}(\mathbf k)
={\cal C}_{Ni}(\mathbf k)
=\frac 14 \int \frac{d k_0}{2\pi}{\rm sign}(k_0) {\rm tr}\left[
{\rm i}{\Sigma\!\!\!/}_{Nii}^{>}(k) {\rm i}S_{Ni}^{<}(k)
-{\rm i}{\Sigma\!\!\!/}_{Nii}^{<}(k) {\rm i}S_{Ni}^{>}(k)
\right]\,.
\end{align}

\subsection{Tree-level collision terms}

The tree-level collision term for the leptons can be straightforwardly 
evaluated by inserting the tree-level propagators into Eq.~(\ref{KB:ell}) 
and the expression (\ref{Sigma:ell}) for the self-energy. Since we 
are interested only in the lepton number density
\begin{equation}
n_l = \int\frac{d^3 k}{(2\pi)^3}\,f_l(\mathbf k),
\end{equation}
we integrate ${\cal C}_\ell(\mathbf k)$ over $\mathbf k$ and 
find
\begin{eqnarray}
\label{C:ell}
&& \int\frac {d^3 k}{(2\pi)^3}\,{\cal C}_\ell(\mathbf k)
= -|Y_i|^2
\int\frac{d^3k}{(2\pi)^3 2|\mathbf k^2|}
\frac{d^3k^{\prime}}{(2\pi)^3 2\sqrt{\mathbf k^{\prime2}+M_i^2}}
\frac{d^3k^{\prime\prime}}{(2\pi)^3 2|\mathbf k^{\prime\prime}|}
\\\nonumber
&& \hspace*{1cm} 
\times \,
(2\pi)^4\delta^4(k^\prime-k-k^{\prime\prime})\,
2k^\prime \cdot k \,\Big\{
(1-f_{Ni}(\mathbf k^\prime))\times[f_\ell(\mathbf k)
f_\phi(\mathbf k^{\prime\prime})-\bar f_\ell(\mathbf k)
\bar f_\phi(\mathbf k^{\prime\prime})]
\\\nonumber
&&  \hspace*{1.4cm} 
-\,f_{Ni}(\mathbf k^\prime)\times[(1-f_\ell(\mathbf k))
(1+f_\phi(\mathbf k^{\prime\prime}))-(1-\bar f_\ell(\mathbf k))
(1+\bar f_\phi(\mathbf k^{\prime\prime}))]
\Big\}\,.
\end{eqnarray}
Due to the fast violation of Higgs number,
we can assume that $f_\phi=\bar f_\phi\equiv f_\phi^{\rm eq}$. For the 
leptons we may assume
$f_\ell-f_\ell^{\rm eq}=-(\bar f_\ell - f_\ell^{\rm eq})$, as justified by 
fast pair-annihilating and -creating interactions among the leptons. 
Expanding to linear order in the deviations from equilibrium, the lepton collision term
simplifies to
\begin{align}\label{coll:ell:lin}
\int\frac {d^3 k}{(2\pi)^3}\,{\cal C}_\ell(\mathbf k)
=&
-|Y_i|^2
\int
\frac{d^3k}{(2\pi)^3 2|\mathbf k|}
\frac{d^3k^{\prime}}{(2\pi)^3 2\sqrt{\mathbf k^{\prime2}+M_i^2}}
\frac{d^3k^{\prime\prime}}{(2\pi)^3 2|\mathbf k^{\prime\prime}|}
\\\notag
&\times
(2\pi)^4\delta^4(k^\prime-k-k^{\prime\prime})\,
2k^\prime \cdot k\,
\left[f^{\rm eq}_\phi(\mathbf k^{\prime\prime})+
f_{Ni}(\mathbf k^\prime)\right]
\times\left[ f_\ell(\mathbf k)-\bar f_\ell(\mathbf k)\right]\,.
\end{align}
This term describes the washout of a lepton asymmetry through inverse decays. 
In a similar way we obtain the tree-level collision terms for 
the neutrinos by inserting the tree-level propagators into Eq.~(\ref{KB:N}) 
and the expression (\ref{Sigma:ell}) for the self-energy. For the 
neutrinos, we shall need the phase-space distributions, so we need the 
unintegrated collision term, for which we find 
\begin{align}
\label{C_N}
{\cal C}_{Ni}(\mathbf k)
=&\frac14 g_w |Y_i|^2
\frac{1}{2\sqrt{{\mathbf k}^2+M_i^2}}
\int\frac{d^3k^\prime}{(2\pi)^3 2|\mathbf k^\prime|}
\frac{d^3k^{\prime\prime}}{(2\pi)^3 2|\mathbf k^{\prime\prime}|}
\,(2\pi)^4\delta^4(k-k^\prime-k^{\prime\prime})
\\\notag
\times
{\rm tr}\Big[&
{k\!\!\!/}^\prime
(k\!\!\!/+M_i)
\left\{
- f_{Ni}(\mathbf k) [1-f_\ell(\mathbf k^\prime)]
[1+f_\phi(\mathbf k^{\prime\prime})]
+ [1-f_{Ni}(\mathbf k)] f_\ell(\mathbf k^\prime)
f_\phi(\mathbf k^{\prime\prime})
\right\}
\\\notag
+&
{k\!\!\!/}^\prime
(k\!\!\!/+M_i)
\left\{
-f_{Ni}(\mathbf k) [1-\bar f_\ell(\mathbf k^\prime)]
[1+\bar f_\phi(\mathbf k^{\prime\prime})]
+ [1-f_{Ni}(\mathbf k)] \bar f_\ell(\mathbf k^\prime)
\bar f_\phi(\mathbf k^{\prime\prime})
\right\}
\Big]
\,.
\end{align}
It is useful to check that when setting the Higgs and lepton 
distribution functions to 
zero, this reproduces the negative total decay rate of $N_i$,
\begin{align}
{\cal C}_{Ni}^{{f,\bar f}_{\ell,\phi}=0}(\mathbf k)
=&
-\frac{g_w}{16\pi} |Y_i^2|\frac{M_i^2}{\sqrt{\mathbf k^2+M_i^2}}
f_{Ni}(\mathbf k) \,.
\end{align}
Next, introducing 
\begin{equation}
\delta f_{Ni}=f_{Ni}-f_{Ni}^{\rm eq},
\end{equation}
and expanding to linear order around equilibrium distributions 
the Majorana collision term simplifies to 
\begin{align}
\label{coll:N:lin}
{\cal C}_{Ni}(\mathbf k)
=&- g_w |Y_i|^2
\frac{1}{2\sqrt{\mathbf k^2+M_i^2}}
\int
\frac{d^3k^\prime}{(2\pi)^3 2|\mathbf k^\prime|}
\frac{d^3k^{\prime\prime}}{(2\pi)^3 2|\mathbf k^{\prime\prime}|}
(2\pi)^4\delta^4(k-k^\prime-k^{\prime\prime})
\\\notag
&\times
2k\cdot k^\prime \,\delta f_{Ni}(\mathbf k) 
[1-f^{\rm eq}_\ell(\mathbf k^\prime)+
f^{\rm eq}_\phi(\mathbf k^{\prime\prime})]\,.
\end{align}
Note that $\delta f_{Ni}$ is not necessarily small in some realistic 
scenarios for leptogenesis, {\it e.g.} for weak washout. 

It is easy to see from Eqs~(\ref{coll:ell:lin}), (\ref{coll:N:lin}) 
that in thermal equilibrium the collision terms vanish. Alternatively, 
one may either by directly use the KMS relations~(\ref{KMS},\ref{KMS:Sigma})  in Eqs.~(\ref{KB:ell},\ref{KB:N}) or explicitly insert the equilibrium 
distributions into Eqs.~(\ref{C:ell},\ref{C_N}). This means 
that the equilibrium state is a static (time-independent) solution to 
the Kadanoff-Baym equations.

\subsection{Expansion of the Universe}

Before proceeding we comment on how the effects of particle dilution 
due to the expansion of the Universe can be systematically incorporated 
into our equations. So far we have formulated the Kadanoff-Baym 
equation on the Minkowski background. Neglecting Planck-scale 
suppressed corrections the only effect of replacing the Minkowski 
background by the Friedmann-Robertson-Walker one is that the particle
modes are red-shifted. 

This can be conveniently implemented by adopting conformal coordinates, 
which are defined by the metric
\begin{align}
\notag
g_{\mu\nu}=a^2(\eta)\,{\rm diag }(1,-1,-1,-1)\,,
\end{align}
where $\eta=x^0$ denotes the conformal time and $a(\eta)$ the scale factor.
Then the Lagrangian~(\ref{Lagrangian}) describes the dynamics in the 
expanding background, provided that we have redefined the fields such 
that their kinetic terms are canonically normalized,
that we replace $M_i\to a M_i$ and that we assume that the Higgs field 
has no coupling to the scalar curvature $R$. Consequently all 
expressions given earlier in this Section and in Sections~\ref{section:wf} 
and~\ref{section:vertex} below remain valid in the expanding Universe 
if momenta $\mathbf k$ are understood as comoving momenta 
$\mathbf k_{\rm com}$, temperature $T$ (and $\beta=1/T$) as comoving 
temperature $T_{\rm com}$ (and $\beta_{\rm com}$), 
time $t$ as conformal time $\eta$, $f(\mathbf k_{\rm com})$ and 
$n$ as comoving phase-space 
distributions and particle number densities, respectively, and 
if finally masses $M_i$ are replaced by $a M_i$, wherever they 
appear.

For the scale factor in the radiation-dominated Universe we use the 
relation
\begin{align}
a(\eta)=a_{\rm R}\eta\,.
\end{align}
Using the Friedmann equation and the free energy, we can relate this to the
temperature as
\begin{equation}
\label{tcom}
T=\frac{T_{\rm com}}{a(\eta)}=
\frac{1}{a(\eta)}\sqrt{\frac{a_{\rm R} m_{\rm Pl}}{2}}
\left(\frac{45}{g_*\pi^3}\right)^{1/4}\,,
\end{equation}
where $g_*$ denotes the number of relativistic degrees of freedom 
and $ m_{\rm Pl} = 1/\sqrt{G_N} = 1.22 \cdot 10^{19}\,\mbox{GeV}$ 
the Planck mass. The variable $z=M_1/T$ that will be used in 
Section~\ref{section:numerics} is therefore related to $\eta$ by 
a constant proportionality factor.

Kinetic equations are often formulated in terms of physical momenta and 
physical time. In order to make contact, we can replace in our 
equations the comoving momenta by physical 
momenta $k_{\rm ph}=k_{\rm com}/a$ and the comoving 
temperature by the physical temperature $T_{\rm ph}=T_{\rm com}/a$. 
Doing so and dividing by a factor 
of $a$ for convenience, the right-hand sides ({\it i.e.} the 
collision terms), simply take the same form
as given earlier in this Section, where now all momenta are to be
understood as physical.
The left-hand sides change according to
\begin{align}
\frac{1}{a(\eta)} \frac{d}{d \eta} f(\mathbf{k}_{\rm com})
&= \frac{\partial}{\partial t} f(\mathbf{k_{\rm ph}})
+\left(\frac{\partial}{\partial|\mathbf k_{\rm ph}|}f(\mathbf{k_{\rm ph}})
\right) \frac{\partial|\mathbf k_{\rm ph}|}{\partial t}
\\\notag
&=\frac{\partial}{\partial t} f(\mathbf{k_{\rm ph}})
-H |\mathbf k_{\rm ph}|
\frac{\partial}{\partial|\mathbf k_{\rm ph}|}f(\mathbf{k_{\rm ph}})
\,,
\end{align}
where $f$ may stand for $f_N$ or $f_\ell-\bar f_\ell$.
This formula applies to the treatment of the densities 
of right-handed neutrinos, that do not necessarily maintain kinetic 
equilibrium during leptogenesis, implying
that one cannot substitute for $f_{Ni}$ a Fermi-Dirac distribution with a 
(pseudo-)chemical potential.
For the lepton charge, it is suitable to consider the integrated version
\begin{align}
\frac{1}{a(\eta)}\int\frac{d^3 k_{\rm ph}}{(2\pi)^3}
\,\frac{d}{d \eta} f(\mathbf{k}_{\rm com})
=\frac{\partial}{\partial t}n_{\rm ph}+3 H n_{\rm ph}
\end{align}
of the above equation. This is just the familiar cosmological dilution 
law, where $n_{\rm ph}$ denotes a physical number density and $H$ the 
Hubble parameter with respect to time $t$. However, in 
Section~\ref{section:numerics} we will solve the kinetic equations 
directly in conformal coordinates.

\subsection{Boltzmann equations}

We now write down explicitly the Boltzmann equations that describe
the time evolution of the distribution function $f_{N1}$ and of the lepton asymmetry
in conformal coordinates:
\begin{align}
\frac{d}{d\eta}f_{N1}(\mathbf k_{\rm com}) &= D(\mathbf k_{\rm com})\,,\label{eq:neutrino} \\
\frac{d}{d\eta}(n_\ell-\bar n_\ell) &= W + S \,.\label{eq:asym}
\end{align}
Here $D(\mathbf k_{\rm com})$ denotes the contributions (\ref{coll:N:lin})
from decays and inverse decays of $N_1$, while $W$ denotes the washout term 
(\ref{coll:ell:lin}). $S=S^{\rm wf}+S^{\rm v}$ is the source term 
for the lepton asymmetry, which we decompose 
into a wave-function contribution $S^{\rm wf}$ and a vertex contribution
$S^{\rm v}$. Note that here and below, unless indicated otherwise,
$n_\ell$ refers to the comoving particle number density.

Performing the substitution of variables pertinent to conformal 
coordinates as discussed above in Eq.~(\ref{C_N}), we obtain:
\begin{align}
\label{kin:eq:N1}
D(\mathbf k_{\rm com})
=&2 g_w |Y_1|^2 \frac{1}{2\sqrt{\mathbf k_{\rm com}^2+a^2(\eta) M_1^2}}
\int
\frac{d^3 k_{\rm com}^{\prime}}{(2\pi)^3 2|\mathbf k_{\rm com}^{\prime}|}
\frac{d^3 k_{\rm com}^{\prime\prime}}{(2\pi)^3 2|\mathbf k_{\rm com}^{\prime\prime}|}
\\\notag
&\times
(2\pi)^4\delta^4(k_{\rm com}-k_{\rm com}^\prime-k_{\rm com}^{\prime\prime})
k_{\rm com}\cdot k_{\rm com}^\prime
\\\notag
&\times
\Big[
-f_{N1}(\mathbf k_{\rm com})
\left(
1-f_\ell\left(\mathbf k_{\rm com}\right) 
+f_\phi\left(\mathbf k_{\rm com}^{\prime\prime}\right)
\right)
+f_\ell\left(\mathbf k_{\rm com}^\prime\right)
f_\phi\left(\mathbf k_{\rm com}^{\prime\prime}\right)
\Big]
\,.
\end{align}
In the evaluation of the tree-level collision term we may set 
$f_\ell$ and $f_\phi$ equal to the equilibrium distributions 
given in Eq.~(\ref{eqdists}) (but expressed in terms of comoving 
momentum and temperature), which allows us to evaluate the 
phase-space integrals. It proves useful to define the quantity
\begin{align}
\label{CP:vec:wf}
\Sigma_N^\mu(k)=g_w\int\frac{d^3p}{(2\pi)^3 2|\mathbf p|}\frac{d^3q}{(2\pi)^3 2|\mathbf q|}
(2\pi)^4 \delta^4(k-p-q)\,p^\mu
\left(
1-f_\ell^{\rm eq}(\mathbf p)+f_\phi^{\rm eq}(\mathbf q)
\right)\,,
\end{align}
which we will calculate analytically in Section~\ref{section:numerics}. 
In terms of this we find for the decay and inverse decay term
\begin{align}
\label{f_N:mode:eq}
D(\mathbf k_{\rm com})=-2 |Y_1|^2\,
\frac{k_{{\rm com}\,\mu}}{2k_{{\rm com}\,0}}\,
\Sigma^\mu_{N}(k_{\rm com})
\,\Big[f_{N1}(\mathbf k_{\rm com})-f^{\rm eq}_{N1}(\mathbf k_{\rm com})
\Big]
\,,
\end{align}
where $k_{{\rm com}\,0}=\sqrt{\mathbf k_{\rm com}^2+a_{\rm R}^2\eta^2 M_1^2}$
and 
\begin{align}
f^{\rm eq}_{N1}(\mathbf k_{\rm com})
=\frac{1}{{\rm e}^{k_{{\rm com}0}/T_{\rm com}}+1}\,.
\end{align}
To calculate the washout term (\ref{coll:ell:lin}) we approximate 
$f_{Ni}$ by the equilibrium distribution $f_{Ni}^{\rm eq}$ and 
assume that the leptons are in kinetic equilibrium so that
$f_\ell$ can be parameterized in terms of a chemical potential. 
Expanding to linear order in the chemical potential, the 
latter assumption yields the relation 
\begin{equation}
f_\ell(\mathbf k)-\bar f_\ell(\mathbf k)
 = (n_l-\bar n_l) \times 
\frac{12\beta^3 e^{\beta |{\mathbf k}|}}
{\left(e^{\beta |{\mathbf k}|}+1\right)^2}\,.
\end{equation}
We then obtain for the wash-out term
\begin{align}
\label{Cl:reduced}
W = \int \frac{d^3 k}{(2\pi)^3}\,
{\cal C}_\ell(\mathbf k)
=-\frac{3|Y_1|^2}{8\pi^3}\frac{a^2 M_1^2}{T_{\rm com}}\,(n_\ell-\bar n_\ell)
\int\limits_0^\infty dx \,\frac{{\rm e}^x}{({\rm e}^x+1)^2}
\ln\left(
\frac{{\rm e}^{\frac{a^2 M_1^2}{4T_{\rm com}^2x}+x}+1}
{{\rm e}^{\frac{a^2 M_1^2}{4T_{\rm com}^2x}+x}-{\rm e}^x}
\right)\,.
\end{align}
The substitution $f_{Ni} \to f_{Ni}^{\rm eq}$, which was made 
to take analytical integrations further,
is strictly speaking only a correct approximation when
$\delta f_{Ni}\ll f_{Ni}^{\rm eq}$. As in scenarios
of strong washout or in the transition region from weak to strong washout
most of the asymmetry
is eventually produced when $N_1$ is close to equilibrium, the error from
assuming an equilibrium abundance of $N_1$ in the washout term at all times
should be small in the final result. (In fact, when  assuming 
Maxwell statistics~\cite{Davidson:2008bu,Giudice:2003jh,Buchmuller:2004nz} 
the absence of  Pauli blocking factors in Eq.~(\ref{C:ell}) 
implies an even more drastic approximation, since then 
$f_{Ni}({\mathbf k^\prime})\to 0$ in Eq.~(\ref{coll:ell:lin}).) 

As is well known, the source term $S$ for the lepton asymmetry 
in Eq.~(\ref{eq:asym}) only 
arises in fourth order in the couplings
$Y_i$. Deriving this source term including all quantum statistical 
factors is the subject of the following two sections. 

\section{\boldmath 
Self-Energy Contribution to the $\,CP$ Asymmetry}
\label{section:wf}

A key ingredient to baryogenesis calculations in the conventional
framework is the subtraction of RIS. If this were omitted, a baryon asymmetry
would result in thermal equilibrium in contradiction with the
CPT theorem and the unitarity of scattering matrices. In this Section we
show how the RIS subtraction arises within the CTP framework for 
leptogenesis and derive
the $CP$ asymmetry from the wave-function correction in the presence of 
finite number densities.

\subsection{\boldmath 
Wave-Function Correction to ${\cal C}_{\ell}$}

The aim is the calculation of the wave-function correction to the 
lepton collision term (\ref{KB:ell}),
\begin{align}
\label{clwave}
{\cal C}^{\rm wf}_{\ell}({\mathbf k})=&
\int\frac{dk_0}{2\pi}
{\rm tr}\left[
{\rm i}{\Sigma\!\!\!/}_\ell^{{\rm wf}>}(k)P_{\rm L}{\rm i}S_{\ell}^<(k)
-{\rm i}{\Sigma\!\!\!/}_\ell^{{\rm wf}<}(k)P_{\rm L}{\rm i}S_{\ell}^>(k)
\right]\,,
\end{align}
where the correction to the lepton self-energy is
\begin{align}
\label{Sigma:ell:wf}
{\rm i}{\Sigma\!\!\!/}^{{\rm wf}<,>}_\ell(k)
=\int\frac{d^4k^\prime}{(2\pi)^4}\frac{d^4k^{\prime\prime}}{(2\pi)^4}
(2\pi)^4\delta^4(k-k^\prime-k^{\prime\prime})
Y_i^*Y_j
P_{\rm R}
{\rm i}S^{{\rm wf}<,>}_{Nij}(k^\prime)
P_{\rm L}
{\rm i}\Delta_\phi^{>,<}(-k^{\prime\prime})\,.
\end{align}
In contrast to Eq.~(\ref{C:ell}) 
the propagator ${\rm i}S^{{\rm wf}<,>}_{Nij}$ 
in this expression is not the tree propagator, but the propagator including 
a one-loop self-energy correction. Hence, we must also allow for off-diagonal 
terms and obtain for the wave-function corrections to
the neutrino Wightman functions:
\begin{subequations}
\begin{align}
-{\rm i}S_{Nij}^{{\rm wf}>}(k^\prime)&=
{\rm i}S_{Ni}^>{\rm i}{\Sigma\!\!\!/}_{Nij}^T {\rm i}S_{Nj}^T
-{\rm i}S_{Ni}^{\bar T}{\rm i}{\Sigma\!\!\!/}_{Nij}^> {\rm i}S_{Nj}^T
-{\rm i}S_{Ni}^>{\rm i}{\Sigma\!\!\!/}_{Nij}^< {\rm i}S_{Nj}^>
+{\rm i}S_{Ni}^{\bar T}{\rm i}{\Sigma\!\!\!/}_{Nij}^{\bar T} {\rm i}S_{Nj}^>
\Big|_{k^\prime}
\,,
\\
-{\rm i}S_{Nij}^{{\rm wf}<}(k^\prime)&=
{\rm i}S_{Ni}^<{\rm i}{\Sigma\!\!\!/}_{Nij}^{\bar T} {\rm i}S_{Nj}^{\bar T}
-{\rm i}S_{Ni}^{T}{\rm i}{\Sigma\!\!\!/}_{Nij}^< {\rm i}S_{Nj}^{\bar T}
-{\rm i}S_{Ni}^<{\rm i}{\Sigma\!\!\!/}_{Nij}^> {\rm i}S_{Nj}^<
+{\rm i}S_{Ni}^{T}{\rm i}{\Sigma\!\!\!/}_{Nij}^{T} {\rm i}S_{Nj}^<
\Big|_{k^\prime}
\,,
\end{align}
\end{subequations}
The different signs are a consequence of the Feynman rules on the CTP, 
and all functions on the right-hand side carry momentum argument 
$k^\prime$ as indicated after the bar.

In order to reduce the number of terms that we need to write,
we assume that $M_2\gg M_1$, such that
the density of $N_2$ can be neglected at the time of leptogenesis. In that
case, the off-diagonal components reduce to
\begin{subequations}
\begin{align}\label{sn12}
-{\rm i}S_{N12}^{{\rm wf}>}&=
{\rm i}S_{N1}^>{\rm i}{\Sigma\!\!\!/}_{N12}^T {\rm i}S_{N2}^T
-{\rm i}S_{N1}^{\bar T}{\rm i}{\Sigma\!\!\!/}_{N12}^> {\rm i}S_{N2}^T\,,\\
-{\rm i}S_{N12}^{{\rm wf}<}&=
{\rm i}S_{N1}^<{\rm i}{\Sigma\!\!\!/}_{N12}^{\bar T} {\rm i}S_{N2}^{\bar T}
-{\rm i}S_{N1}^{T}{\rm i}{\Sigma\!\!\!/}_{N12}^< {\rm i}S_{N2}^{\bar T}
\,,\\
-{\rm i}S_{N21}^{{\rm wf}>}&=
-{\rm i}S_{N2}^{\bar T}{\rm i}{\Sigma\!\!\!/}_{N21}^> {\rm i}S_{N1}^T
+{\rm i}S_{N2}^{\bar T}{\rm i}{\Sigma\!\!\!/}_{N21}^{\bar T} {\rm i}S_{N1}^>\,,\\
-{\rm i}S_{N21}^{{\rm wf}<}&=
-{\rm i}S_{N2}^{T}{\rm i}{\Sigma\!\!\!/}_{N21}^< {\rm i}S_{N1}^{\bar T}
+{\rm i}S_{N2}^{T}{\rm i}{\Sigma\!\!\!/}_{N21}^{T} {\rm i}S_{N1}^<
\,.
\end{align}
\end{subequations}
As long as $|M_2-M_1|\gg\Gamma_{N1},\Gamma_{N2}$, where $\Gamma_{Ni}$ are 
the total decay rates of $N_i$, we can straightforwardly extend the 
present analysis to situations where $M_2\gg M_1$ does not hold by 
simply adding the corresponding contributions that arise from a 
non-vanishing density of $N_2$, i.e. the asymmetry produced in $N_2$ decays. 
The reason for this is that not both of the neutrinos in the wave-function
diagram can be on-shell simultaneously, provided their mass terms are not
degenerate. Because of this we do not further simplify 
${\rm i}S_{N2}^{T,\bar T}$ to $-{\rm i}/M_2$ at this point.

\subsection{KMS and Real Intermediate State Subtraction}

It follows from Eq.~(\ref{clwave}) that ${\cal C}_\ell^{\rm wf}$ 
vanishes provided both ${\Sigma\!\!\!/}_\ell^{{\rm wf}<,>}$ and 
$S_\ell^{<,>}$ satisfy the KMS relation. The correction
${\Sigma\!\!\!/}_\ell^{{\rm wf}<,>}$ in turn satisfies KMS provided
it also holds for ${\rm i}S_{Nij}^{{\rm wf}<,>}$, 
see Eq.~(\ref{Sigma:ell:wf}). It is instructive to
show this property,
\begin{align}
\label{KMS:SNij}
{\rm i}S_{Nij}^{{\rm wf}>}(k)
=-{\rm e}^{\beta k_0} {\rm i}S_{Nij}^{{\rm wf}<}(k)
\end{align}
explicitly, because it illustrates how the RIS subtraction is realised
within the CTP framework.

To show the validity of the KMS relation for the wave-function 
correction~(\ref{KMS:SNij}) explicitly, we define
\begin{align}
{\cal K}_{ij}={\rm i} S^{{\rm wf}>}_{Nij}(k) + 
e^{\beta k_0} {\rm i} S^{{\rm wf}<}_{Nij}(k).
\end{align}
The claim is ${\cal K}_{ij}=0$ when substituting equilibrium distributions.
We show this on the example ${\cal K}_{12}$, from which it will be 
clear how to generalize to all components of ${\cal K}_{ij}$.
Applying KMS to ${\Sigma\!\!\!/}_{N12}^{<,>}$ and $S_{N1}^{<,>}$, we find
\begin{align}
\label{K12}
{\cal K}_{12}=-{\rm i}S_{N1}^> {\rm i}
\left(
{\Sigma\!\!\!/}_{N12}^T+{\Sigma\!\!\!/}_{N12}^{\bar T}
\right){\rm i}S_{N2}^T
+{\rm i}S_{N1}^T {\rm i}{\Sigma\!\!\!/}_{N12}^>{\rm i}S_{N2}^T
+{\rm i}S_{N1}^{\bar T} {\rm i}{\Sigma\!\!\!/}_{N12}^>{\rm i}S_{N2}^T\,.
\end{align}
In addition, we have used that $S_{N2}$ is evaluated here only when $N_2$ is
off-shell, such that we can set $S_{N2}^{\bar T}=-S_{N2}^T$.
To further simplify this expression, we make use of the following identities:
\begin{align}
S_{N{\rm i}}^T + S_{N\rm i}^{\bar{T}} &= S^>_{N{\rm i}}+ S^<_{N{\rm i}}\,, \\
	{\Sigma\!\!\!/}_{Nij}^T+{\Sigma\!\!\!/}_{Nij}^{\bar T}
&={\Sigma\!\!\!/}_{Nij}^>+{\Sigma\!\!\!/}_{Nij}^<\,.
\end{align}
Note that the first identity, applied to the heavy neutrino $N_{\rm 2}$, 
recovers the property $S^T_{N2} = - S^{\bar{T}}_{N2}$, when evaluated for 
off-shell momenta. It follows that
\begin{align}
{\cal K}_{12}=-{\rm i}S_{N1}^> {\rm i}
\left(
{\Sigma\!\!\!/}_{N12}^<+{\Sigma\!\!\!/}_{N12}^>
\right){\rm i}S_{N2}^T
+
{\rm i}(S_{N1}^<+{\rm i}S_{N1}^>) {\rm i} {\Sigma\!\!\!/}_{N12}^>
{\rm i}S_{N2}^T=0\,,
\end{align}
where we have once more applied the KMS relation to obtain the last equality.
An immediate consequence of this result is that no asymmetry is 
produced in thermal equilibrium, as required.

\begin{figure}[t!]
\begin{center}
\epsfig{file=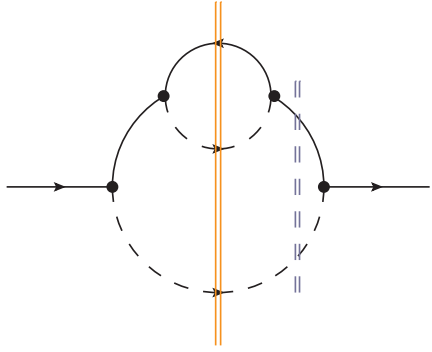,width=6cm}
\end{center}
\caption{
\label{fig:wf}
Diagrammatic representation of the lepton-number violating
contribution to ${\Sigma\!\!\!/}_\ell^{{\rm wf}<,>}$. The solid lines
with arrows represent the lepton $\ell$, the solid lines without arrows the
neutrinos $N_{1,2}$ and the dashed lines with arrows the Higgs boson $\phi$.
The solid (light grey/orange) double line represents the cut that corresponds
to the subtracted real intermediate states when finite density corrections to
$CP$ violation are neglected. The dashed (dark-grey/blue) double 
line represents the cut that corresponds to decays and inverse decays 
when finite density corrections are neglected.}
\end{figure}

Besides verifying the 
KMS relation for $S_{Nij}^{<,>}$ and consequently also for
${\Sigma\!\!\!/}_{Nij}^{<,>}$ and ${\Sigma\!\!\!/}_\ell^{<,>}$, the
above calculation shows how the RIS subtraction emerges in the CTP framework.
To see this, we neglect the finite density corrections in
$\Sigma_{\ell}^{<,>}$ implying that we set
$f_\ell=\bar f_\ell=f_\phi=\bar f_\phi\equiv 0$ in the propagators of 
the $T$ and $\bar T$ type. Then, in order to see which
process a certain contribution in the collision term corresponds to,
we  cut through the propagators of the $<,>$ type, while
leaving the propagators of the $T,\bar T$ type intact.
We can then identify the terms in Eq.~(\ref{K12}) that involve
${\Sigma\!\!\!/}_{Nij}^{T,\bar T}$ and $S_{N1}^{<,>}$ with
decays and inverse decays. In the Feynman diagram representation of
${\Sigma\!\!\!/}_\ell$, that is given in Figure~\ref{fig:wf}, this
corresponds to the cut indicated by the dashed double line, which gives
the interference term between the tree-level decay amplitudes
and the wave-function corrections to the decay amplitudes.
The terms in  Eq.~(\ref{K12}) that involve
${\Sigma\!\!\!/}_{Nij}^{<,>}$ and $S_{N1}^{T,\bar T}$ correspond to
scatterings. Since $S_{N1}^{T}$ and $S_{N1}^{\bar T}$ are added together,
precisely the pole contribution, when $N_1$ is on-shell, is isolated
from the scatterings, which takes the effect of RIS subtraction.
In the Feynman diagram in Figure~\ref{fig:wf} this
corresponds to the cut indicated by the solid double lines. It
represents the the squared amplitudes of the scatterings of leptons and 
Higgs bosons through $N_1$ and $N_2$ as well as the interference terms 
between these, which are
crucial  for the cancellation of $CP$-violating effects in equilibrium.
Within the full CTP framework, ${\Sigma\!\!\!/}_{Nij}^{T,\bar T}$
and the $T,\bar T$-type propagators also acquire
contributions from real particles in the plasma, such that the clear 
separation in decay and scattering processes through the cuts in 
Figure~\ref{fig:wf} no longer persists.

In order to see the role of the violation of time reversal symmetry, 
we note that for the vanishing of ${\cal K}_{ij}$, we make use of the 
KMS relation in the form of
\begin{align}
{\rm i}S_{Ni}^>(k){\rm i}S_\ell^<(k^\prime)
{\rm i}\Delta_\phi^<(k^{\prime\prime})
=
{\rm i}S_{Ni}^<(k){\rm i}S_\ell^>(k^\prime)
{\rm i}\Delta_\phi^>(k^{\prime\prime})
\end{align} 
for momenta satisfying $k=k^\prime+k^{\prime\prime}$. 
This equation is  valid not only in thermal equilibrium, 
but it is also another way of expressing that the process 
$N_i\to\ell\phi$ and its inverse  $\ell\phi\to N_i$
occur at the same rate, and that the same holds true for
$N_i\to\bar\ell\phi^*$ and $\bar\ell\phi^*\to N_i$. 
Hence, the above discussion does
not only show that no $CP$ asymmetry is generated in equilibrium, but more
generally that no asymmetry is generated in situations that are symmetric
with respect to time reversal, as it is of course expected by the 
$CPT$ theorem.

\subsection{\boldmath $CP$ Source}

We now derive the $CP$-violating source within the lepton collision
term, that arises due to deviations of $N_1$ from equilibrium and
a violation of the KMS relations. 
First, we note that because of Eq.~(\ref{prop:N:expl}) we can identify
\begin{align}
{\rm i}\delta S_{N1}^>={\rm i}\delta S_{N1}^<
={\rm i}\delta S_{N1}^T={\rm i}\delta S_{N1}^{\bar T}
\equiv {\rm i}\delta S_{N1}\,,
\end{align}
where ${\rm i}\delta S_{N1}^{>,<,T,\bar T}$ denotes the difference between
the non-equilibrium neutrino propagators and their equilibrium versions.
Using this the $CP$-source collision term (\ref{clwave}) reads
\begin{eqnarray}
{\cal C}_{\ell}^{\rm wf}(\mathbf k)
&=&\int\frac{d k_0}{2\pi}
\frac{d^4k^\prime}{(2\pi)^4}\frac{d^4k^{\prime\prime}}{(2\pi)^4}
(2\pi)^4\delta^4(k-k^\prime-k^{\prime\prime})
\\\nonumber
&&\hskip-1cm
\times \,(-1)\,\Bigg\{\left[
Y_1^* Y_2 {\rm i}\delta S_{N1}
 {\rm i}\left({\Sigma\!\!\!/}_{N12}^T-{\Sigma\!\!\!/}_{N12}^>\right)
{\rm i} S_{N2}^T
-Y_1 Y_2^*{\rm i}S_{N2}^T
\left({\Sigma\!\!\!/}_{N21}^{\bar T}-{\Sigma\!\!\!/}_{N21}^>\right)
{\rm i}\delta S_{N1}
\right]_{k^\prime}
\nonumber\\
&& \hspace*{1cm}\times\,
{\rm i}\Delta_\phi^<(-k^{\prime\prime}){\rm i}S_{\ell}^<(k)
\nonumber \\[0.2cm]\notag
&&\hskip0.3cm+\,
\left[
Y_1^* Y_2 {\rm i}\delta S_{N1}
 {\rm i}\left({\Sigma\!\!\!/}_{N12}^{\bar T}-{\Sigma\!\!\!/}_{N12}^<\right)
{\rm i} S_{N2}^T
-Y_1 Y_2^*{\rm i}S_{N2}^T
\left({\Sigma\!\!\!/}_{N21}^{T}-{\Sigma\!\!\!/}_{N21}^<\right)
{\rm i}\delta S_{N1}
\right]_{k^\prime}
\nonumber\\
&& \hspace*{1cm}\times\,
{\rm i}\Delta_\phi^>(-k^{\prime\prime}){\rm i}S_{\ell}^>(k)
\Bigg\}
\,.\nonumber
\end{eqnarray}
In this expression, the terms involving the dispersive parts
${\rm }{\Sigma\!\!\!/}_{Nij}^{{\rm disp}T,\bar T}$ of 
${\rm }{\Sigma\!\!\!/}_{Nij}^{T,\bar T}$ cancel.
To see this, first note that
${\rm i}{\Sigma\!\!\!/}_{Nij}^{{\rm disp}T}=
-{\rm i}{\Sigma\!\!\!/}_{Nij}^{{\rm disp}{\bar T}}$.
Then, since $\delta S_{N1}(k_0,\mathbf k)=\delta S_{N1}(-k_0,{\mathbf k}) $ 
and since for equilibrium distributions
${\rm i}\Delta_\phi^<(k_0^\prime,\mathbf k^\prime)
{\rm i}S_{\ell}^<(k_0^\prime,\mathbf k^\prime)
={\rm i}\Delta_\phi^>(-k_0^\prime,\mathbf k^\prime)
{\rm i}S_{\ell}^>(-k_0^\prime,\mathbf k^\prime)$,
the cancellation occurs when performing the integrals.
We can therefore make the replacements
\begin{align}
\label{ret:to:spec}
{\Sigma\!\!\!/}^{T,\bar T}_{ij}
-{\Sigma\!\!\!/}^{>}_{ij}
\to\frac 12\left(
{\Sigma\!\!\!/}^{<}_{ij}
-{\Sigma\!\!\!/}^{>}_{ij}
\right)\,,
\quad
{\Sigma\!\!\!/}^{T,\bar T}_{ij}
-{\Sigma\!\!\!/}^{<}_{ij}
\to
-\frac 12\left(
{\Sigma\!\!\!/}^{<}_{ij}
-{\Sigma\!\!\!/}^{>}_{ij}
\right)
\,,
\end{align}
and write the wave-function correction to the lepton collision integral as
\begin{align}
\label{C:wf:coll}
\int\frac{d^3k}{(2\pi)^3}{\cal C}_\ell^{\rm wf}(\mathbf k)
=&
\int\frac{d^4k}{(2\pi)^4}\frac{d^4k^\prime}{(2\pi)^4}
\frac{d^4k^{\prime\prime}}{(2\pi)^4}\,
(2\pi)^4\delta^4(k-k^\prime-k^{\prime\prime})
\\\notag
&\hskip-2cm
\times
{\rm tr}
\Bigg[
\left(
-Y_1^* Y_2 {\rm i}\delta S_{N1} \frac{\rm i}{2}\left({\Sigma\!\!\!/}^<_{N12}
-{\Sigma\!\!\!/}^>_{N12}\right)
{\rm i}S_{N2}^T
+Y_1 Y_2^* {\rm i}S_{N2}^T \frac{\rm i}{2}\left({\Sigma\!\!\!/}^<_{N21}
-{\Sigma\!\!\!/}^>_{N21}\right)
{\rm i}\delta S_{N1}
\right)_{k^\prime}
\\\notag
&\hspace*{-1cm} \times
\Big(
{\rm i}\Delta^<_\phi(-k^{\prime\prime})
{\rm i}S^<_{\ell}(k)
-
{\rm i}\Delta^>_\phi(-k^{\prime\prime})
{\rm i}S^>_{\ell}(k)
\Big)
\Bigg]\,.
\end{align}
The spectral self-energy is
\begin{align}
{\rm i}{\Sigma\!\!\!/}_{Nij}^<(k)-{\rm i}{\Sigma\!\!\!/}_{Nij}^>(k)
=&-\int\frac{d^3k^\prime}{(2\pi)^3 2|\mathbf k^\prime|^2}
\frac{d^3k^{\prime\prime}}{(2\pi)^3 2|\mathbf k^{\prime\prime}|^2}
(2\pi)^4\delta^4(k-k^{\prime}-k^{\prime\prime})
\\\notag
&\times{\rm sign}(k_0) \left[
Y_i Y_j^* {k\!\!\!/}^\prime P_{\rm R}
+Y_i^* Y_j {k\!\!\!/}^\prime P_{\rm L}
\right]\times
[1-f^{\rm eq}_\ell(\mathbf{k}^\prime)+
f^{\rm eq}_\phi(\mathbf{k}^{\prime\prime})]
\,.
\end{align}
In further simplifying ${\cal C}_\ell^{\rm wf}$, we obtain another factor
of ${\rm sign}(k_0)$ from
\begin{align}
&{\rm i}\Delta^<_\phi(-k^{\prime\prime})
{\rm i}S^<_{\ell}(k)
- {\rm i}\Delta^>_\phi(-k^{\prime\prime})
{\rm i}S^>_{\ell}(k)
\big|_{{\rm sign}(k_0)=-{\rm sign}(k_0^{\prime\prime})}
\\\notag
&=-P_{\rm L}k\!\!\!/P_{\rm R}\,
(2\pi)^2
\delta(k^2)\delta({k^{\prime\prime}}^2)
{\rm sign}(k_0)\left(1-f^{\rm eq}_\ell(\mathbf k)+
f^{\rm eq}_\phi(\mathbf{k}^{\prime\prime})\right)\,.
\end{align}
Using this result
in Eq.~(\ref{C:wf:coll}), the tree-level propagators, and substituting
after performing the Dirac trace
the definition~(\ref{CP:vec:wf}),
we obtain
our final result for the $CP$-violating wave-function correction 
to the lepton collision term:
\begin{align}
\label{coll:wf:final}
S^{\rm wf}=&\int\frac{d^3k}{(2\pi)^3}{\cal C}_\ell^{\rm wf}(\mathbf k)
\\\notag
=&\int
\frac{d^3k}{(2\pi)^3 2|\mathbf k|}
\frac{d^3k^\prime}{(2\pi)^3 2\sqrt{{\mathbf k^\prime}^2+M_1^2}}
\frac{d^3k^{\prime\prime}}{(2\pi)^3 2|\mathbf k^{\prime\prime}|}
(2\pi)^4 \delta^4(k^\prime-k-k^{\prime\prime})
\\\notag
&\times
\frac{M_1M_2}{{k^\prime}^2-M_2^2}
2{\rm Im}[Y_1^2 {Y_2^*}^2]
2 k_\mu \Sigma^\mu_N(k^\prime) \delta f_N(\mathbf k^\prime)
\left(1-f_\ell^{\rm eq}(\mathbf k)+
f_\phi^{\rm eq}(\mathbf k^{\prime\prime})\right)
\\
=&\,4\,{\rm Im}[Y_1^2 {Y_2^*}^2]\,\frac{M_1M_2}{M_1^2-M_2^2}
\int\frac{d^3k^\prime}{(2\pi)^3 2\sqrt{{\mathbf k^\prime}^2+M_1^2}}
\,\delta f_N(\mathbf k^\prime)\,
\frac{\Sigma_{N\mu}(\mathbf k^\prime)\Sigma_N^{\mu}(\mathbf k^\prime)}{g_w}
\,.\notag
\end{align}
Note that at zero temperature, we find
\begin{align}\
\label{Sigma_mu:vac}
\Sigma_N^\mu(k)\Big|_{T=0}=g_w \frac{k^\mu}{16\pi}.
\end{align}
Then, by comparison with the tree-level lepton collision term~(\ref{C:ell}), 
we reproduce the zero-temperature result for the decay asymmetry
\begin{align}
\label{CP:par:vac}
\varepsilon^{\rm wf}_{T=0}=\frac{\Gamma_{N1\to\ell H}-
\Gamma_{N1\to \bar\ell H^\dagger}}{\Gamma_{N1\to\ell H}+
\Gamma_{N1\to \bar\ell H^\dagger}}
=\frac{1}{8\pi}\frac{{\rm Im}[Y_1^2 {Y_2^*}^2]}
{|Y_1|^2}\frac{M_1 M_2}{M_1^2 -M_2^2}
\,,
\end{align}
where $\Gamma$ denotes the decay rate of the process indicated in the 
subscript. On the other hand, if finite-density effects are kept for 
the initial and final states but not in the self-energy loop, then 
in the final expression of Eq.~(\ref{coll:wf:final}) one substitutes 
\begin{equation}
\label{nofdinloop}
\frac{\Sigma_{N\mu}(\mathbf k^\prime)\Sigma_N^{\mu}(\mathbf k^\prime)}{g_w} 
\to \frac{k^{\prime\,\mu}\Sigma_{N\mu}(\mathbf k^\prime)}{16\pi}\,.
\end{equation}

\section{\boldmath 
Vertex Contribution to the $\,CP$ Asymmetry}
\label{section:vertex}

\subsection{\boldmath 
Vertex Correction to ${\cal C}_\ell$}

The vertex correction contribution to the $CP$ asymmetry arises 
from the two-loop self-energy diagram for the lepton propagator 
(see Figure~\ref{fig:vertex}):
\begin{align}
{\rm i}{\Sigma\!\!\!/}^{{\rm v}ab}_{\ell}(k)
=& - cd \,{Y_i^*}^2 {Y_j}^2
\int \frac{d^4p}{(2\pi)^4}\frac{d^4q}{(2\pi)^4}\,
P_{\rm R}
{\rm i}S_{Ni}^{ac}(-p)
C\left[P_{\rm L}{\rm i}S_\ell^{dc}(p+k+q)P_{\rm R}\right]^t C^\dagger
\\\notag
& \times
{\rm i}S_{Nj}^{db}(-q)
P_{\rm L}
{\rm i}\Delta^{da}_\phi(-p-k)
{\rm i}\Delta^{bc}_\phi(-q-k)
\,.
\end{align}
Here, the transposition $t$ acts on the Dirac indices, and $i$, $j$  
and the CTP indices $c,d=\pm$ are summed over. The charge conjugation arises
as a consequence of lepton number violation and the reversed flow
of the internal lepton in the self-energy diagram.  

We again consider
the case where $M_2\gg M_1$ in order to reduce the number of terms 
to be considered by one half. Furthermore, we neglect the 
diagonal $(i=j)$ contributions to the self-energy since they do not 
contribute to the $CP$ asymmetry. This can be easily
verified along the lines of the calculation presented in this Section. 
With these simplifications, we obtain
\begin{eqnarray}
\label{sigma:ell:greater}
&& {\rm i}{\Sigma\!\!\!/}_\ell^{{\rm v}>}(k) =
-\int \frac{d^4p}{(2\pi)^4}\frac{d^4q}{(2\pi)^4}
\\\notag
&& \times\, \Big\{
{Y_1^*}^2 {Y_2}^2
\Big[
{\rm i}S_{N1}^{>}(-p)
C\left[P_{\rm L}{\rm i}S_{\ell}^{T}(p+k+q)P_{\rm R}\right]^t C^\dagger
{\rm i}S_{N2}^{T}(-q)
{\rm i}\Delta_{\phi}^{<}(-p-k){\rm i}\Delta_{\phi}^{T}(-q-k)
\\\notag
&& \hspace*{1cm}-
{\rm i}S_{N1}^{\bar T}(-p)
C\left[P_{\rm L} {\rm i}S_{\ell}^{<}(p+k+q)P_{\rm R}\right]^t C^\dagger
{\rm i}S_{N2}^{T}(-q)
{\rm i}\Delta_{\phi}^{<}(-p-k){\rm i}\Delta_{\phi}^{<}(-q-k)
\Big]
\\\notag
&& + \, {Y_1}^2 {Y_2^*}^2
\Big[
- {\rm i}S_{N2}^{\bar T}(-p)
C\left[P_{\rm L}{\rm i}S_{\ell}^{<}(p+k+q)P_{\rm R}\right]^t C^\dagger
{\rm i}S_{N1}^{T}(-q)
{\rm i}\Delta_{\phi}^{<}(-p-k){\rm i}\Delta_{\phi}^{<}(-q-k)
\\\notag
&& \hspace*{1cm} +\,
{\rm i}S_{N2}^{\bar T}(-p)
C\,[P_{\rm L} {\rm i}S_{\ell}^{\bar T}(p+k+q)P_{\rm R}]^t C^\dagger
{\rm i}S_{N1}^{>}(-q)
{\rm i}\Delta_{\phi}^{\bar T}(-p-k){\rm i}\Delta_{\phi}^{<}(-q-k)
\Big]\Big\}.
\end{eqnarray}
The expression for ${\rm i}{\Sigma\!\!\!/}_\ell^{{\rm v}<}(k)$ is 
obtained through interchange
of the CTP indices $T\leftrightarrow \bar{T}$ and $<\,\leftrightarrow \,>$. 

\subsection{KMS and Real Intermediate State Subtraction}

Before proceeding with the calculation of the $CP$ source, we again
show that this contribution leads to a vanishing asymmetry in equilibrium,
which in particular means that the KMS relation,
${\rm i}{\Sigma\!\!\!/}_\ell^{{\rm v}>}(k)+
{\rm e}^{\beta k_0}{\rm i}{\Sigma\!\!\!/}_\ell^{{\rm v}<}(k)=0$, must hold when
equilibrium distributions are inserted.
To see explicitly how this arises, we consider the contributions
$\propto {Y_1^*}^2{Y_2}^2$. First, we define
\begin{align}
{\rm i}{\Sigma\!\!\!/}_\ell^{{\rm v}>}(k)
+{\rm e}^{\beta k_0}{\rm i}{\Sigma\!\!\!/}_\ell^{{\rm v}<}(k)
=\int \frac{d^4p}{(2\pi)^4}\frac{d^4q}{(2\pi)^4}
{\cal J}(k,p,q)
\end{align}
and note that by virtue of the KMS relation
\begin{eqnarray}
{\cal J}(k,p,q)
& = & {Y_1^*}^2 {Y_2}^2 
\left[
{\rm i}S_{N1}^{T}(-p)+{\rm i}S_{N1}^{\bar T}(-p)
\right]
C\left[P_{\rm L} {\rm i}S_{\ell}^{<}(p+k+q)P_{\rm R}\right]^t C^\dagger
{\rm i}S_{N2}^{T}(-q)
\qquad \\\notag
&& \times\,
{\rm i}\Delta_{\phi}^{<}(-p-k){\rm i}\Delta_{\phi}^{<}(-q-k)
\\\notag
&-& {Y_1^*}^2 {Y_2}^2 
{\rm i}S_{N1}^{>}(-p)
C\Big[
P_{\rm L}{\rm i}S_{\ell}^{T}(p+k+q)P_{\rm R}\Delta_{\phi}^{T}(-q-k)
\\\notag&&
+\,P_{\rm L}{\rm i}S_{\ell}^{\bar T}(p+k+q)P_{\rm R}
\Delta_{\phi}^{\bar T}(-q-k)
\Big]^t C^\dagger{\rm i}S_{N2}^{T}(-q)
{\rm i}\Delta_{\phi}^{<}(-p-k)\,.
\end{eqnarray}
Further simplification is achieved when deriving from the explicit
expressions~(\ref{prop:ell:expl}) and~(\ref{prop:phi:expl}) that
for ${\rm sign}(r_0)={\rm sign}(s_0)$,
\begin{align}
{\rm i}S_\ell^T(r){\rm i}\Delta_\phi^T(s)+{\rm i}S_\ell^{\bar T}(r){\rm i}\Delta_\phi^{\bar T}(s)
={\rm i}S_\ell^<(r){\rm i}\Delta_\phi^<(s)+{\rm i}S_\ell^{>}(r){\rm i}\Delta_\phi^{>}(s)
\,.
\end{align}
We find
\begin{eqnarray}
{\cal J}(k,p,q)& = &{Y_1^*}^2 {Y_2}^2 
\left[
{\rm i}S_{N1}^{<}(-p)+{\rm i}S_{N1}^{>}(-p)
\right]
C\left[P_{\rm L} {\rm i}S_{\ell}^{<}(p+k+q)P_{\rm R}\right]^t C^\dagger
{\rm i}S_{N2}^{T}(-q)\qquad 
\\\notag&&\times
{\rm i}\Delta_{\phi}^{<}(-p-k){\rm i}\Delta_{\phi}^{<}(-q-k)
\\\notag
&-&{Y_1^*}^2 {Y_2}^2 
{\rm i}S_{N1}^{>}(-p)
C\Big[
P_{\rm L}{\rm i}S_{\ell}^{<}(p+k+q)P_{\rm R}\Delta_{\phi}^{<}(-q-k)
\\\notag&&
+P_{\rm L}{\rm i}S_{\ell}^{>}(p+k+q)P_{\rm R}\Delta_{\phi}^{>}(-q-k)
\Big]^t C^\dagger
{\rm i}S_{N2}^{T}(-q){\rm i}\Delta_{\phi}^{<}(-p-k)
\\\notag
&=&0\,,
\end{eqnarray}
where for the last equality, we have again used the KMS relation. 
Likewise, one can show
the vanishing of the terms $\propto{Y_1}^2{Y_2^*}^2$ in equilibrium. 

\begin{figure}[t]
\begin{center}
\epsfig{file=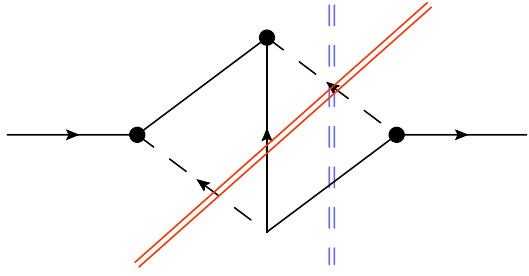,width=7cm}
\end{center}
\caption{
\label{fig:vertex}
Diagrammatic representation of ${\Sigma\!\!\!/}_\ell^{{\rm v}<,>}$. The solid lines
with arrows represent the lepton $\ell$, the solid lines without arrows the
neutrinos $N_{1,2}$ and the dashed lines with arrows the Higgs boson $\phi$.
The solid (light-grey/orange) double line represents the cut that corresponds
to the subtracted real intermediate states when finite density corrections to
$CP$ violation are neglected. The dashed (dark-grey/blue) double line 
represents the cut that corresponds to decays and inverse decays 
when finite density corrections are neglected.
}
\end{figure}

Again, we can relate this verification of the KMS relation to the 
standard procedure of RIS subtraction.
As in the wave-function contribution, at zero temperature the $<,>$ type 
Green functions correspond to cut propagators. The terms in 
(\ref{sigma:ell:greater})  involving $S_{N1}^{<,>}$ and 
$\Delta_\phi^{<,>}$ thus correspond to the cut indicated by
the dashed double line in Figure \ref{fig:vertex} that represents the 
interference between the tree-level and one-loop contribution to the $N_1$ 
decay amplitude. The terms involving $S_{N1}^{T,\bar{T}}$ on the other 
hand correspond to $N_1$ mediated scatterings, as indicated by the solid 
cut line in Figure \ref{fig:vertex}. 
Note that here the cut corresponds to the interference between  $s$-channel 
and  $t$-channel diagrams, while the central cuts through the wave-function 
correction yield the squares of the $s$-channel and $t$-channel 
contributions to the scatterings.
This complementarity is crucial for the separate
cancellation of the asymmetries
from the wave function and the vertex correction in equilibrium.

\subsection{\boldmath $CP$ Source}

We now extract the $CP$ source in the lepton collision term using the 
same simplifications
as in the calculation of the wave-function correction. In particular, 
the deviation of $N_1$ 
from equilibrium is parametrised again by $\delta S_{N1}$. We obtain
\begin{align}
\label{Cv:long}
&\hskip-5.0cm
{\cal C}_\ell^{\rm v}(\mathbf k)=
\int\frac{d k_0}{2\pi}
{\rm tr}
\left[
{\rm i}{\Sigma\!\!\!/}_\ell^{{\rm v}>}(k){\rm i}S_\ell^{<}(k)
-{\rm i}{\Sigma\!\!\!/}_\ell^{{\rm v}<}(k){\rm i}S_\ell^>(k)
\right]\\\notag
&\hspace*{-4.8cm} 
=\int \frac{d k_0}{2\pi}\frac{d^4p}{(2\pi)^4}\frac{d^4q}{(2\pi)^4}
\times\,{\rm tr}\Big[
\Big\{-
{Y_1^*}^2{Y_2}^2
{\rm i}\delta S_{N1}(-p)
C\big[
{\rm i}S_\ell^T(p+k+q){\rm i}\Delta_\phi^T(-q-k)
\\\notag
-{\rm i}S_\ell^<(p+k+q)&{\rm i}\Delta_\phi^<(-q-k)
\big]^t
C^\dagger
{\rm i} S_{N2}^T(-q){\rm i}\Delta_\phi^<(-p-k)
\\\notag
+{Y_1}^2{Y_2^*}^2
{\rm i} S_{N2}^T(-p)
C \big[&
{\rm i}S_\ell^{\bar T}(p+k+q){\rm i}\Delta_\phi^{\bar T}(-q-k)
\\\notag
-{\rm i}S_\ell^<(p+k+q)&{\rm i}\Delta_\phi^<(-q-k)
\big]^t C^\dagger
{\rm i} \delta S_{N1}^T(-q){\rm i}\Delta_\phi^<(-p-k)
\Big\}\,
{\rm i}S_\ell^<(k)
\\\notag& 
\hspace*{-3.8cm} - \Big\{\! -{Y_1}^2{Y_2^*}^2
{\rm i} S_{N2}^T(-p)
C \big[
{\rm i}S_\ell^{T}(p+k+q){\rm i}\Delta_\phi^{T}(-q-k)
\\\notag
-{\rm i}S_\ell^>(p+k+q)&{\rm i}\Delta_\phi^>(-q-k)
\big]^t C^\dagger
{\rm i} \delta S_{N1}^T(-q){\rm i}\Delta_\phi^>(-p-k)
\\\notag&
\hspace*{-3.8cm}+{Y_1^*}^2{Y_2}^2
{\rm i}\delta S_{N1}(-p)
C\big[
{\rm i}S_\ell^{\bar T}(p+k+q){\rm i}\Delta_\phi^{\bar T}(-q-k)
\\\notag
-{\rm i}S_\ell^>(p+k+q)&{\rm i}\Delta_\phi^>(-q-k)
\big]^t
C^\dagger
{\rm i} S_{N2}^T(-q){\rm i}\Delta_\phi^>(-p-k)
\Big\}
\,{\rm i}S_\ell^>(k)
\Big]
\,,
\end{align}
where ${\rm i} S_{N2}^{\bar{T}} =-{\rm i} S_{N2}^T$  for off-shell 
neutrinos $N_2$ has been used. Inspection of the Dirac
structure reveals that $S_{N1,2}$ only contribute through terms proportional 
to $M_{1,2}$, which is familiar from the standard calculation of the decay
asymmetry~\cite{Fukugita:1986hr}. Finally, similar to the
replacement~(\ref{ret:to:spec}) we can substitute here
\begin{align}
S_\ell^{T,\bar T}\Delta_\phi^{T,\bar T}-S_\ell^>\Delta_\phi^>
&\to
+ \frac12
\left(
S_\ell^<\Delta_\phi^<-S_\ell^>\Delta_\phi^>
\right)\,,
\\\notag
S_\ell^{T,\bar T}\Delta_\phi^{T,\bar T}-S_\ell^<\Delta_\phi^<
&\to
-\frac12
\left(
S_\ell^<\Delta_\phi^<-S_\ell^>\Delta_\phi^>
\right)
\end{align}
under the integrals.

The expression~(\ref{Cv:long})
closely resembles the result for the wave-function contribution, except that
${\rm i}S_{N2}^T$ now depends on the loop momentum $q$. 
In the limit where $M_2\gg M_1$, the $N_2$ propagator can be expanded in 
powers of $\tilde{k}^2/M_2^2$ where $\tilde{k} = (p,k)$, showing that in 
this limit the vertex correction is just one half of the self-energy 
contribution, where the factor one half 
arises since there is no summation over SU(2) indices within the loop. 

Inserting the explicit expressions for the lepton and Higgs boson Green 
functions, we obtain our final result for the vertex correction to the 
$CP$ asymmetry. To this end we introduce
\begin{align}
\label{GammaV}
\Gamma_\mu(k,p^{\prime\prime},M_2)
=&\int
\frac{d^3 k^\prime}{(2\pi)^3 2|\mathbf k^\prime|}
\frac{d^3 k^{\prime\prime}}{(2\pi)^3 2|\mathbf k^{\prime\prime}|}
(2\pi)^4\delta^4(k-k^\prime-k^{\prime\prime})\,
k_\mu^\prime\,\frac{M_1 M_2}{(k^\prime-p^{\prime\prime})^2-M_2^2}
\\\notag
&\times
\left[
1-f_\ell(\mathbf k^\prime)+f_\phi(\mathbf k^{\prime\prime})
\right]\,.
\end{align}
This has to be integrated over the initial and final state phase space 
in the collision term, such that we additionally define
\begin{align}
V(k,M_2)
=&\int\frac{d^3 p^\prime}{(2\pi)^3 2|\mathbf p^\prime|}
\frac{d^3 p^{\prime\prime}}{(2\pi)^3 2|\mathbf p^{\prime\prime}|}
(2\pi)^4\delta^4(k-p^\prime-p^{\prime\prime})\,
{p^\prime}^\mu\,\Gamma_\mu(k,p^{\prime\prime},M_2)
\\\notag
&\times
\left[
1-f_\ell(\mathbf p^\prime)+f_\phi(\mathbf p^{\prime\prime})
\right]\,.
\end{align}
In terms of these quantities, the lepton collision term can be expressed as
\begin{align}
\label{coll:vertex:final}
S^{\rm v}=\int\frac{d^3 p^\prime}{(2\pi)^3}\,
{\cal C}_\ell^{\rm v}(\mathbf p^\prime)
=4\,{\rm Im}[Y_1^2 {Y_2^*}^2]\int \frac{d^3 k}{(2\pi)^3}\,
\delta f_{N1}(\mathbf k) \,V(k,M_2)
\,.
\end{align}

\subsection{Source term in the strongly hierarchical 
limit}

For the strongly hierarchical case, $M_2\gg M_1$, which we will focus on 
for the numerical analysis, the vertex contribution is simply one half 
of the self energy contribution. To see this, we note that in the limit 
$M_2\gg M_1$, 
\begin{equation}
\frac{M_1 M_2}{(k^\prime-p^{\prime\prime})^2-M_2^2} 
\to -\frac{M_1}{M_2}
\end{equation}
in Eq.~(\ref{GammaV}), which implies 
\begin{equation}
\Gamma_\mu(k,p^{\prime\prime},M_2) = 
-\frac{M_1}{M_2}\,\frac{\Sigma_\mu(k)}{g_W}
\qquad \mbox{and} \qquad
V(k,M_2) = -\frac{M_1}{M_2} 
\frac{
\Sigma_{N\mu}(\mathbf k^\prime)\Sigma_N^{\mu}(\mathbf k^\prime)
}{g_w^2}.
\end{equation}
Inserting this into (\ref{coll:vertex:final}) and adding the 
wave-function contribution (\ref{coll:wf:final}) the source 
term for the lepton asymmetry in Eq.~(\ref{eq:asym}) is given by
\begin{align}
\label{vertex:source}
S_{M_2\gg M_1}  = \frac{3}{2}\times\,4\,{\rm Im}[Y_1^2 {Y_2^*}^2] 
\left(-\frac{M_1}{M_2}\right) \int
\frac{d^3k^\prime}{(2\pi)^3 2\sqrt{{\mathbf k^\prime}^2+M_1^2}}
\, \delta f_N(\mathbf k^\prime)\,
\frac{
\Sigma_{N\mu}(\mathbf k^\prime)\Sigma_N^{\mu}(\mathbf k^\prime)
}
{g_w}\,,
\end{align}
with $\Sigma_{N\mu}(\mathbf{k}')$ given in Eq.~(\ref{CP:vec:wf}).

\section{Numerical Estimates of Finite Density Effects}
\label{section:numerics}

In this Section we solve the kinetic equation for the final lepton asymmetry 
with $CP$-violating collision terms derived in 
Sections~\ref{Section:Tree:Level}~--~\ref{section:vertex}. We
compare our results with all finite density effects included to 
the case where these effects are included in the initial and final 
state phase-space distributions, but not in the quantum corrections. 

\subsection{\boldmath Effective $\,CP$-violating parameter}

To begin with, however, we present a rough estimate for the total magnitude 
of $CP$ violation in the lepton collision term. This estimate is provided
through an effective $CP$-violating parameter, which we define as:
\begin{align}
\varepsilon_{\rm eff} \equiv 
-\frac{\frac{d}{d\eta}(n_{\ell}-\bar{n}_{\ell}) - W}{\frac{d}{d\eta}n_{N1}}\,,
\end{align}
where $n_{\ell,N1} = \int\frac{d^3k}{(2\pi)^3}f_{\ell,N1}(\mathbf k)$ are 
the comoving number densities of lepton and neutrino. The tree-level (washout) 
contribution $W$, see Eq.~(\ref{Cl:reduced}), is subtracted from this 
definition, so that the numerator involves only the $CP$ source, which is 
comprised of self-energy and vertex type corrections to the lepton collision 
term, given by Eqs.~(\ref{coll:wf:final},~\ref{coll:vertex:final}). 
The $CP$-violating parameter thus also breaks into the corresponding two 
parts. For the self-energy part we get by using 
Eqs.~(\ref{f_N:mode:eq}), (\ref{coll:wf:final}):
\begin{align}
\label{CP:par:eff}
\varepsilon^{\rm wf}_{\rm eff} = \varepsilon^{\rm wf}_{T=0} \frac{16\pi}{g_w}
\frac{\int d{|{\mathbf k}|} \frac{|{\mathbf k}|^2}{k_0}\,
\Sigma_{N1}^\mu(k)\Sigma_{N1\,\mu}(k) \,\delta f_{N1}(\mathbf  k)}
{\int d{|{\mathbf k}|} \frac{|{\mathbf k}|^2}{k_0} \,
k_\mu \Sigma_{N1}^\mu(k) \,\delta f_{N1}(\mathbf  k)}\,,
\end{align}
where ${\varepsilon^{\rm wf}_{T=0}}$ is the $CP$-violating parameter in 
the vacuum, given by Eq.~(\ref{CP:par:vac}). We see right away, that 
Eq.~(\ref{CP:par:eff}) indeed reduces to ${\varepsilon^{\rm wf}_{T=0}}$ 
in the limit where the finite density effects in the loop are neglected 
in which case one factor of $\Sigma^\mu_N(k)$ in the numerator is 
replaced by $g_w k^\mu/(16\pi)$. We will not compute the vertex correction 
part here in full generality, but we note that in the strongly 
hierarchical case, where $M_2\gg M_1$, the vertex $CP$-source reduces to 
one-half of the self energy $CP$-source, as shown in the previous Section. 
The same is then obviously true for the corresponding $CP$-violating 
parameter $\varepsilon^{\rm v}_{\rm eff}$, so that the ratio 
$\varepsilon_{\rm eff}/\varepsilon_{T=0}$ is the same for the total and 
the wave-function contribution alone.

\begin{figure}[t!]
\begin{center}
\epsfig{file=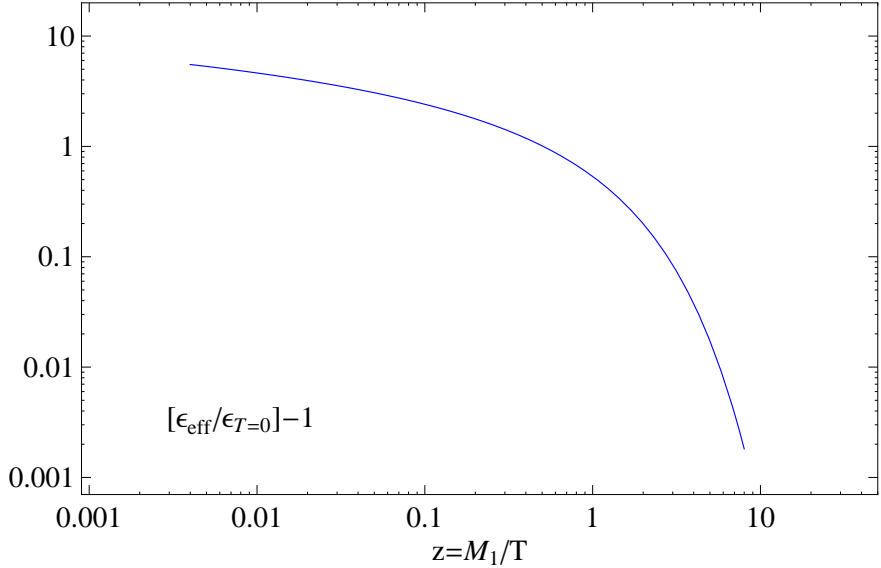,width=12cm}
\end{center}
\caption{
\label{fig:CP:num}
The ratio of the $CP$-violating parameter when taking effects of finite density
in the loop into account over the $CP$-violating parameter at $T=0$.}
\end{figure}

To compute $\varepsilon^{\rm wf}_{\rm eff}$ requires the knowledge of 
the neutrino distribution function $f_{N1}(\mathbf  k)$ for each instant 
of time (or temperature in the expanding Universe), which is obtained by 
solving the tree-level Boltzmann equation (\ref{f_N:mode:eq}). Before 
doing this we make an estimate for $\varepsilon^{\rm wf}_{\rm eff}$ by 
factoring out the nontrivial time-dependence in Eq.~(\ref{CP:par:eff}). 
This can be achieved by assuming that the decaying heavy neutrino $N_1$ 
remains in kinetic equilibrium and making an ansatz with a pseudo-chemical 
potential $\mu_{N1}$: 
\begin{align}
 f_{N1}(\mathbf k) = \frac{1}{\exp[\beta(\sqrt{{\mathbf k}^2 + M_1^2} 
- \mu_{N1})] + 1}\,.
\end{align}
For small deviations from thermal equilibrium, this
can be expanded in linear order of $\mu_{N1}$ as
\begin{align}
 \delta f_{N1}(\mathbf k) &= f_{N1}^{\rm eq}(\mathbf k)
\left(1 - f_{N1}^{\rm eq}(\mathbf k)\right)\frac{\mu_{N1}}{T}\,.
\end{align} 
Using this ansatz the pseudo-chemical potential $\mu_{N1}$ involving the 
time dependence cancels out in the ratio in Eq.~(\ref{CP:par:eff}), and we get
\begin{align}
\label{CP:par:eff:ansatz}
\varepsilon^{\rm wf}_{\rm eff}
=\varepsilon^{\rm wf}_{T=0}  \frac{16\pi}{g_w}
\frac{\int d{|{\mathbf k}|} \frac{|{\mathbf k}|^2}{k_0}\,
\Sigma_{N1}^\mu(k)\Sigma_{N1\,\mu}(k) \,
f_{N1}^{\rm eq}(\mathbf  k)\left(1 - f_{N1}^{\rm eq}(\mathbf  k)\right)}
{\int d{|{\mathbf k}|} \frac{|{\mathbf k}|^2}{k_0}\,
 k_\mu \Sigma_{N1}^\mu(k) \,
f_{N1}^{\rm eq}(\mathbf k)\left(1 - f_{N1}^{\rm eq}(\mathbf  k)\right)}\,.
\end{align}
To evaluate this expression, we see that the phase space integrals 
in $\Sigma_{N1}^\mu(k)$, Eq.~(\ref{CP:vec:wf}), can be computed to give
\begin{subequations}
\label{Sigma:analytic}
\begin{align}
\Sigma_N^0(k) &= \frac{g_w T^2}{8 \pi |{\mathbf k}|}\,
I_1\!\left(\frac{k_0}{T},\frac{|\mathbf k|}{T}\right)\,,
\\
\Sigma^i_N(k) &= \frac{g_w T^2}{8 \pi |{\mathbf k}|}
\left(\frac{k_0}{|{\mathbf k}|}\,I_1\!\left(\frac{k_0}{T},
\frac{|\mathbf k|}{T}\right) - \frac{M_1^2}{2|{\mathbf k}|T}\, 
I_0\!\left(\frac{k_0}{T},\frac{|\mathbf k|}{T}\right) \right)
\frac{k^i}{|{\mathbf k}|}\,,
\end{align}
\end{subequations}
with
\begin{align}
I_n(y_0,y) \equiv \int\limits_{\frac{1}{2}(y_0-y)}^{\frac{1}{2}(y_0+y)} 
dx\;x^n \left(1-\frac{1}{{\rm e}^x+1} + \frac{1}{{\rm e}^{y_0-x}-1}\right)\,,
\end{align}
which have analytical expressions
\begin{align}
I_0(y_0,y) =& \ln\left(\frac{e^{y_0+y}-1}{e^{y_0-y}-1}\right) - y\,,
\\[2mm]
I_1(y_0,y) =& \frac{1}{2}(y_0+y)\ln\left(\frac{1+e^{\frac{y_0+y}{2}}}
{1-e^{\frac{-y_0+y}{2}}}\right) - \frac{1}{2}(y_0-y)
\ln\left(\frac{1+e^{\frac{y_0-y}{2}}}{1-e^{\frac{-y_0-y}{2}}}\right)
\\\notag
&+{\rm Li}_2\left(-e^{\frac{y_0+y}{2}}\right) + 
{\rm Li}_2\left(e^{\frac{-y_0-y}{2}}\right) - 
{\rm Li}_2\left(-e^{\frac{y_0-y}{2}}\right) - 
{\rm Li}_2\left(e^{\frac{-y_0+y}{2}}\right)\,,
\end{align}
where ${\rm Li}_2$ denotes dilogarithm. The remaining one-dimensional 
momentum integrations in Eq.~(\ref{CP:par:eff:ansatz}) can then easily be 
computed numerically. We plot the resulting ratio 
$\varepsilon^{\rm wf}_{\rm eff} / \varepsilon^{\rm wf}_{T=0}$ (equal 
to $\varepsilon_{\rm eff} / \varepsilon_{T=0}$ in the strongly 
hierarchical limit) as a function of $z=M_1/T$ in Figure~\ref{fig:CP:num}. 
We find that the medium effects tend to enhance the effective asymmetry, 
in particular, the Bose enhancement of the Higgs particle dominates over the
Fermi suppression of the lepton.

\subsection{Numerical solution of the Boltzmann equations}

We now scrutinise how the enhancement of the asymmetry is reflected
in the solution of the Boltzmann equations 
(\ref{eq:neutrino}),~(\ref{eq:asym}).
For our numerical analysis we consider the asymmetries in the limit of 
strong hierarchy: $M_2\gg M_1$, when the total $CP$ source in 
Eq.~(\ref{eq:asym}) is given by Eq.~(\ref{vertex:source}). 
The vector $\Sigma_N^\mu$ can be evaluated according to 
Eqs.~(\ref{Sigma:analytic}) when substituting for $T$ 
the ``comoving'' temperature $T_{\rm com}=a(\eta)T$. The washout term 
on the right-hand side of Eq.~(\ref{eq:asym}) has been given in 
Eq.~(\ref{Cl:reduced}).

In the parametric region of strong washout, the neutrino density and
also the lepton asymmetry will eventually become independent of the 
initial conditions, while this is not the case in the weak-washout 
domain. When dropping the assumption that the right-handed neutrinos are in 
equilibrium~\cite{Basboll:2006yx,HahnWoernle:2009qn},
we can integrate the equations for the
comoving modes $\mathbf k_{\rm com}$ of the neutrino number distribution
separately, which is the method
that we pursue here. Thus, to solve the time-evolution equation for 
the lepton asymmetry, we first calculate the out-of-equilibrium 
neutrino distribution by solving Eq.~(\ref{eq:neutrino}) with 
the decay and inverse decay term $D(\mathbf k_{\rm com})$ 
given by Eq.~(\ref{f_N:mode:eq}). This solution is then inserted 
into the source term. The expression~(\ref{f_N:mode:eq}) carries dependence 
on conformal time $\eta$ through the dependence on  
$k_{{\rm com}\,0}=\sqrt{\mathbf k_{\rm com}^2+a_{\rm R}^2\eta^2 M_1^2}$, 
and the dependence of $\Sigma^\mu_{N}(k_{\rm com})$ and
$f^{\rm eq}_{N1}(\mathbf k_{\rm com})$ on $k_{{\rm com}\,0}$. 
Thus a numerical solution of Eq.~(\ref{eq:neutrino}) for the 
out-of-equilibrium neutrino distribution is necessary.

\begin{figure}[t!]
\begin{center}
\epsfig{file=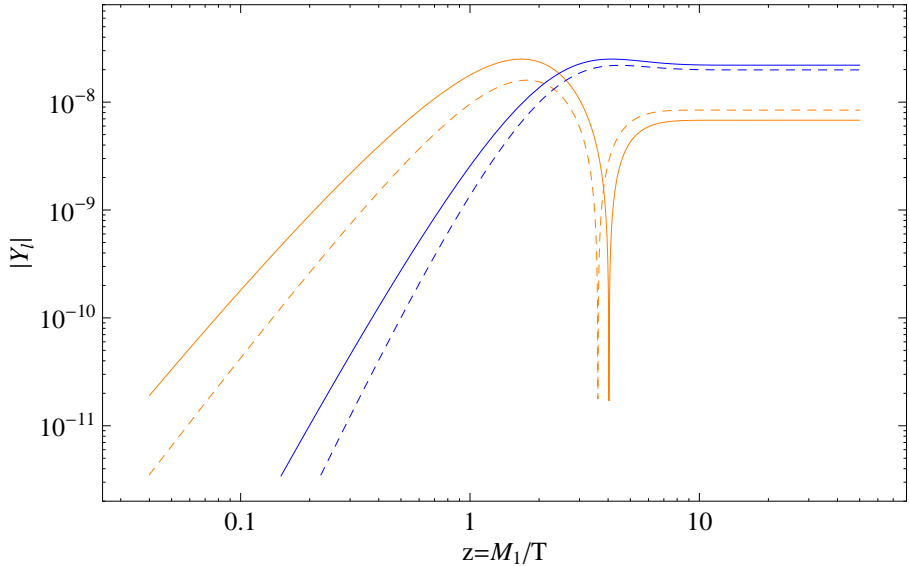,width=12cm}
\end{center}
\caption{
\label{figure:BMann:smallY}
The absolute value of lepton-to-entropy ratio $Y_\ell$ over
$z=M_1/T$. The parameters are
$M_1=10^{13}\,{\rm GeV}$,
$M_2=10^{15}\,{\rm GeV}$, $Y_1=2\times10^{-2}$, $Y_2=10^{-1}$ and
a maximal $CP$ phase. Both, thermal initial conditions (dark grey/blue) and
vanishing initial conditions (light grey/red) for $N_1$ are chosen.
The solid lines correspond to solutions
where finite density corrections in the loop are taken into account,
the dotted lines to solutions where these are omitted.
}
\end{figure}

\begin{figure}[p]
\begin{center}
\epsfig{file=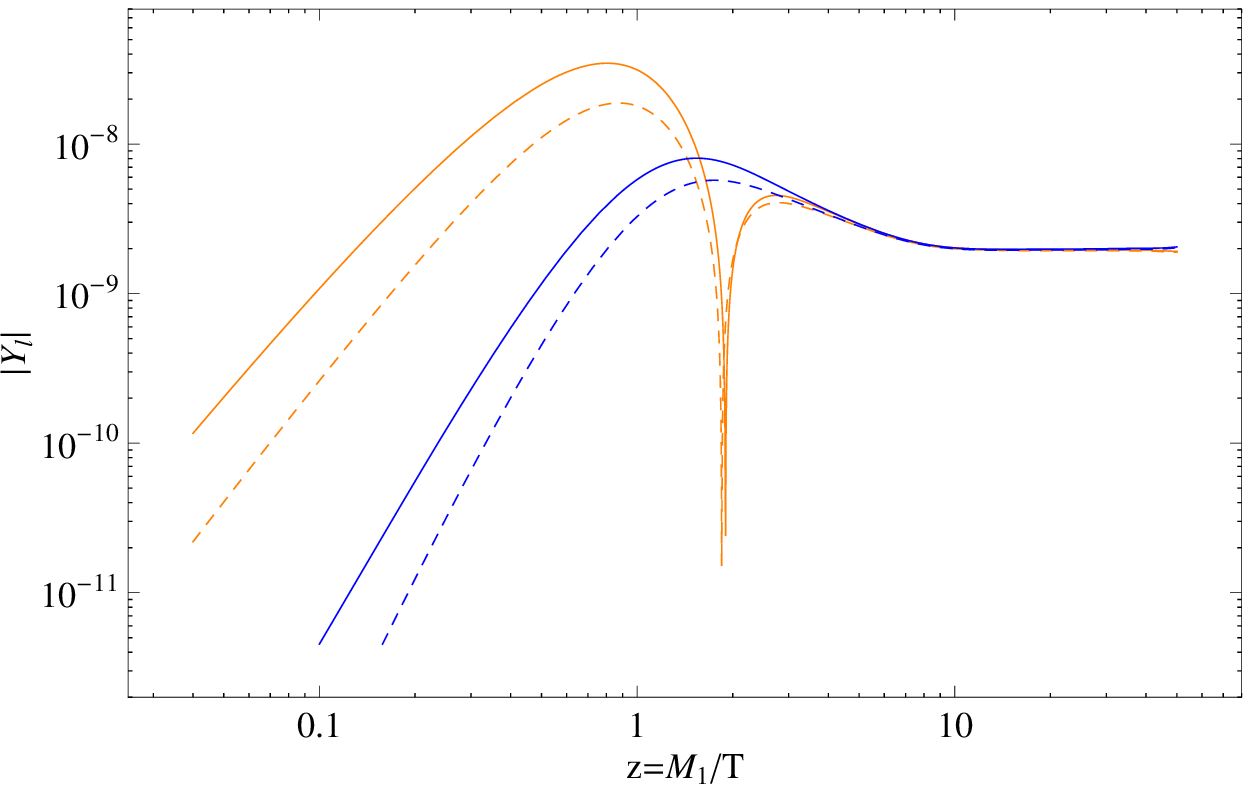,width=12cm}
\vskip1cm
\epsfig{file=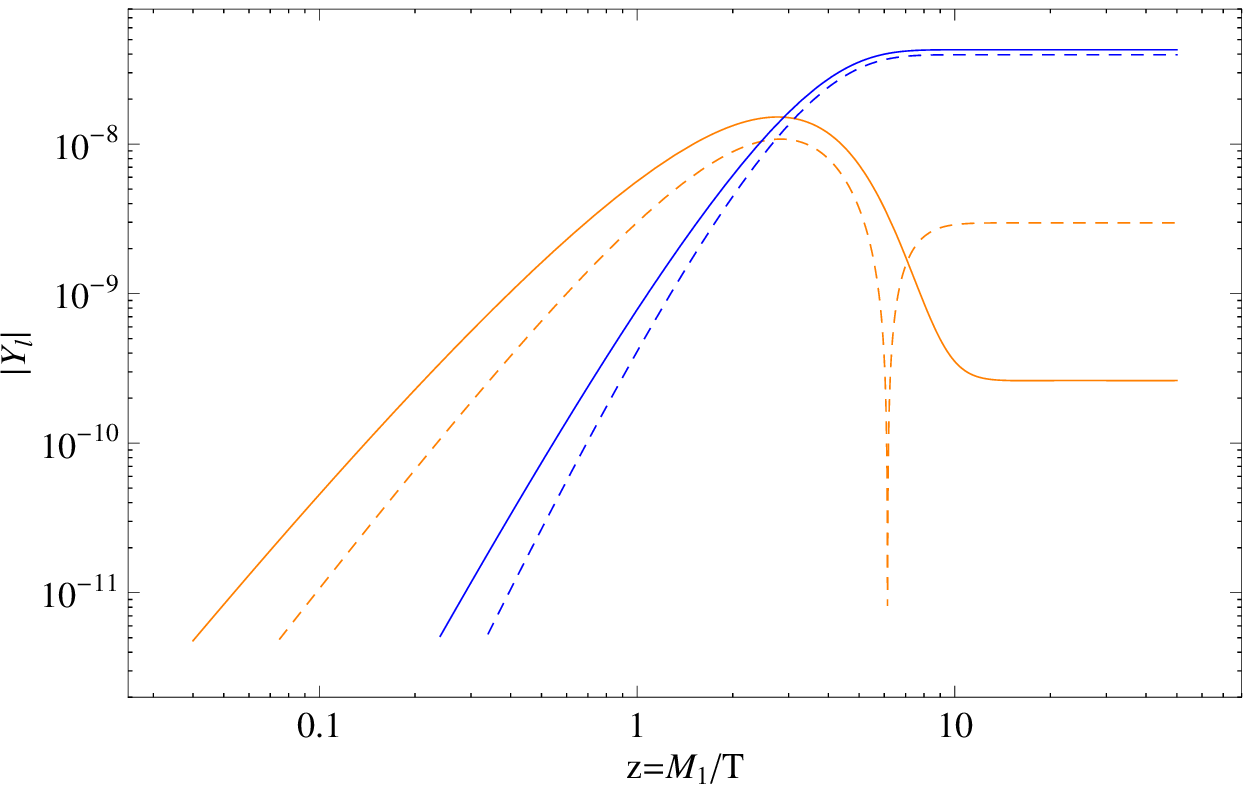,width=12cm}
\end{center}
\caption{
\label{figure:BMann:largeY}
The absolute value of lepton-to-entropy ratio $Y_\ell$ over
$z=M_1/T$. The parameters are
$M_1=10^{13}\,{\rm GeV}$,
$M_2=10^{15}\,{\rm GeV}$, $Y_1=5\times10^{-2}$, $Y_2=10^{-1}$ and
a maximal $CP$ phase in the upper panel (strong washout) and 
 $Y_1=10^{-2}$, $Y_2=10^{-1}$ in the lower panel (weak washout).
Both, thermal initial conditions (dark grey/blue) and
vanishing initial conditions (light grey/red) for $N_1$ are chosen.
The solid lines correspond to solutions
where finite density corrections in the loop are taken into account,
the dotted lines to solutions where these are omitted.
}
\end{figure}

In Figure~\ref{figure:BMann:smallY}, we show solutions to the Boltzmann
equations in the intermediate regime between weak and strong washout. 
What is shown is the asymmetry summed over spin and SU(2) degree of 
freedoms and normalized to the entropy density,
\begin{equation}
Y_\ell=2 g_w \frac{n_{\rm ph, \ell}-\bar n_{\rm ph, \ell}}{s}
= 2 g_w \frac{n_{\rm com, \ell}-\bar n_{\rm com, \ell}}{\frac{2\pi^2}{45} 
g_\star T^3_{\rm com}},
\end{equation}
where the last form is convenient since we calculate directly 
the comoving particle densities. We assume an effective number of
relativistic degrees of freedom according to the Standard Model 
($g_*=106.75$). In the Figure, we provide results 
both with (solid lines) and without (dashed lines) 
the additional finite density enhancement in the 
loops. We choose the parameters $M_1=10^{13}\,{\rm GeV}$,
$M_2=10^{15}\,{\rm GeV}$, $|Y_1|=2\times10^{-2}$, $|Y_2|=10^{-1}$ and
a maximal $CP$ phase such that ${\rm Im}[Y_1^2 {Y_2^*}^2]$ is purely 
imaginary. Besides, we consider two different initial conditions, 
namely vanishing (light grey/red lines) and thermal 
(dark grey/blue lines) initial distributions for $N_1$.
Recall also that the parameter $z=M_1/T$ is proportional to $\eta$.
We observe from the Figure that there is a sizable effect of order 
$(10-20)\%$ from the finite density terms in the loops surviving 
at late times, since the asymmetry is partly produced at early times, 
when $M_1$ is not much larger
than $T$ and quantum statistical corrections are of importance.
This situation no longer persists
at larger coupling, $Y_1=5\times10^{-2}$, corresponding
to strong washout. We display the result in the upper panel of 
Figure~\ref{figure:BMann:largeY}.
Here the final asymmetry is determined at late times
when $M_1\gg T$, such that quantum statistical factors have no sizable impact.

Finally, we consider a scenario of weak washout (here we choose $Y_1=10^{-2}$).
The resulting asymmetries are
displayed in the lower panel of 
Figure~\ref{figure:BMann:largeY}. In the case of
vanishing initial conditions for $N_1$,
we find the impact of the
finite-density effects in the loop in this scenario quite drastic:
When they are taken into account, the asymmetry produced at early times
when the $N_1$-abundance is below its equilibrium value becomes larger,
due to the Bose enhancement of the Higgs particles. This ``wrong-sign'' 
asymmetry is then too large to be cancelled by the ``right-sign'' asymmetry 
produced at later times, when the  $N_1$-abundance is above its 
equilibrium value. Not only the temperature dependence of the
washout, but also the temperature-dependence of the effective $CP$-violating
parameter is crucial for determining the final lepton asymmetry.
This feature does not occur when the loop enhancement is neglected, 
since in that case, the effective $CP$ asymmetry
is temperature independent and the resulting asymmetry is always of
the``right-sign''. 
Therefore, neglecting the finite density effects in the loop
in weak washout scenarios is in general not justifiable and may
under certain circumstances even lead to an
error in the predicted sign of the asymmetry.

\section{Conclusions}
\label{section:conclusions}

We presented a full derivation and numerical solution of the 
kinetic equations describing leptogenesis in the CTP framework.
The results extend existing approaches since
all quantum statistical factors are accounted for systematically  
for the first time in a realistic model
of leptogenesis. We have put particular 
emphasis on establishing a connection to the conventional Boltzmann 
approach and elucidated how RIS subtraction is realised in the CTP 
formalism. The numerical analysis shows that finite density 
corrections do not play a significant role in the strong-washout 
regime while they may have a sizable impact
in weak-washout scenarios.

The detailed analysis of the simple model of leptogenesis presented 
in this work should be seen  as a first step toward applications of 
the CTP formalism to situations where the Boltzmann framework needs 
to be modified. One may think of flavour 
effects~\cite{Endoh:2003mz,Pilaftsis:2005rv,Abada:2006fw,Nardi:2006fx},
the inclusion of additional thermal effects~\cite{Giudice:2003jh} or resonant
leptogenesis~\cite{Flanz:1994yx,Pilaftsis:1997jf,Pilaftsis:2003gt,Pilaftsis:2005rv} in the limit of degenerate right-handed neutrino masses.
Examples of approaches to these more advanced problems using the 
CTP formalism may already be found the 
literature~\cite{De Simone:2007rw,Anisimov:2010aq,Cirigliano:2009yt}.
However, so far there has been no complete
derivation and solution of the equations describing the
standard scenario of leptogenesis within the CTP framework. The present 
work may be viewed as a self-consistent leading-order approximation 
in this framework, which serves as a starting point for investigating 
further corrections and variations.

\subsubsection*{Note added}

While this paper was in writing, Ref.~\cite{Garny:2010nj} appeared 
which also studies modifications of an effective 
$CP$-violating parameter in a realistic leptogenesis model due to 
finite-density effects in loops.

\subsubsection*{Acknowledgements}

\noindent
This work is supported by the Gottfried Wilhelm Leibniz programme 
of the Deutsche Forschungsgemeinschaft and by the Schweizer Nationalfonds.
PS would like to thank RWTH Aachen University for hospitality.

\end{document}